\newcommand{\bra}[1]{\langle #1 \vert}
\newcommand{\ket}[1]{\vert #1 \rangle}
\newcommand{\av}[1]{\langle #1 \rangle}

\documentclass[12pt]{iopart}
\usepackage{iopams}  

\usepackage{dsfont}
\usepackage{graphicx}
\usepackage{appendix}

\begin{document}

\title[Geometrical interpretation of the argument of weak values]{Geometrical interpretation of the argument of weak values of general observables in $N$-level quantum systems}

\author{Lorena Ballesteros Ferraz$^1$, Dominique L Lambert$^2$ and Yves Caudano$^1$}

\address{$^1$Research Unit Lasers and Spectroscopies (UR-LLS), naXys
\& NISM, University of Namur, Rue de Bruxelles 61, B-5000 Namur, Belgium}
\ead{lorena.ballesteros@unamur.be, yves.caudano@unamur.be}

\address{$^2$ ESPHIN \& naXys, 
University of Namur, Rue de Bruxelles 61, B-5000 Namur, Belgium}

\begin{abstract}

Observations in quantum weak measurements are determined by complex numbers called weak values. We present a geometrical interpretation of the argument of weak values of general Hermitian observables in $N$-dimensional quantum systems in terms of geometric phases. We formulate an arbitrary weak value in function of three real vectors on the unit sphere  in $N^2-1$ dimensions, $S^{N^2-2}$. These vectors are linked to the initial and final states, and to the weakly measured observable, respectively. We express pure states in the complex projective space of $N-1$ dimensions, $\mathbb{C}\textrm{P}^{N-1}$, which has a non-trivial representation as a $2N-2$ dimensional submanifold of $S^{N^2-2}$ (a generalization of the Bloch sphere for qudits). The argument of the weak value of a projector on a pure state of an $N$-level quantum system describes a geometric phase associated to the symplectic area of the geodesic triangle spanned by the vectors representing the pre-selected state, the projector and the post-selected state in  $\mathbb{C}\textrm{P}^{N-1}$. We then proceed to show that the argument of the weak value of a general observable is equivalent to the argument of an effective Bargmann invariant. Hence, we extend the geometrical interpretation of projector weak values to weak values of general observables. In particular, we consider the generators of SU($N$) given by the generalized Gell-Mann matrices. Finally, we study in detail the case of the argument of weak values of general observables in two-level systems and we illustrate weak measurements in larger dimensional systems by considering projectors on degenerate subspaces, as well as Hermitian quantum gates.

\end{abstract}

\maketitle

\section{Introduction}

The second revolution of quantum physics promises to improve significantly our technological abilities beyond the current state of the art \cite{Acin2018Qrev2, Laucht2021Qrev2, Deutsch2020QRev2, leontica2021simulating, bian2021nanoscale, cooke2021first}. At the heart of quantum technologies, and of quantum mechanics in general, measurements play a crucial role. Indeed, measurements are the only way to extract information from quantum systems \cite{busch2016quantum, jacobs2014quantum}, be it for achieving practical or fundamental goals. The most common type of measurement is the ideal and non-reversible, i. e. projective, measurement that randomly collapses the system state to one eigenstate of the measured observable. Most often in that case, the details of the interaction between the system and measuring device are overlooked. However, von Neumann developed a model of the interaction between the system and the device that describes their joint evolution during during a quantum measurement  \cite{mello2014vonNeumann}. The stronger the interaction, the more information the measurement provides  but, inevitably as well, the larger the system disturbance \cite{fuchs1996PRAdisturbance, naus2021quantum}.

Quantum weak measurements involve minimal interactions between a measuring device, also called the ancilla, and the system: they arise when the ancilla state couples very weakly with the system state \cite{aharonov1988result, svensson2013pedagogical}. Weakness is defined by the validity of a first-order perturbation expansion of the unitary operator specifying the evolution of the complete quantum state (including both the system itself and the ancilla). When measuring a system observable $\hat{A}$ in the von Neumann scheme, the joint evolution to first order in the coupling strength $g$ is typically given by $\hat{U}\approx\hat{\mathds{1}}-\textrm{i} g \hat{A} \otimes \hat{p}$, where $\hat{p}$ represents the ancilla's canonical momentum. This evolution entangles the system and ancilla states, so that the knowledge of the ancilla state at the end of the measurement process provides information on the system observable. However, in a weak measurement, the identity component of the Taylor development of the unitary operator dominates the evolution of the complete system. Hence, the final state remains nearly identical to the initial one. Advantageously, weak measurements do not destroy completely the system initial state $\ket{\psi_i}$. They have thus many theoretical and practical applications \cite{jacobs2014quantum, svensson2013pedagogical, dressel2014RMPweakvalue}.

In what is typically called a weak measurement in the literature, the measurement weakness is considered in conjunction with the post-selection of the system state. Post-selection is equivalent to constraining the final state to a particular state $\ket{\psi_f}$. Operationnally, it is performed by a strong, projective, final, measurement of the system consecutively to the weak interaction. The projection impacts the ancilla state, shifting its wavefunction centroid by a quantity that is related to what is called the weak value of the observable $\hat{A}$ for the initial and final states \cite{dressel2014RMPweakvalue}:
\begin{equation}
        \label{eq:weak_value_definition}
        A_w=\frac{\bra{\psi_f}\hat{A}\ket{\psi_i}}{\langle\psi_f\ket{\psi_i}}=\frac{\Tr{(\hat{\Pi}_f\hat{A}\hat{\Pi}_i)}}{\Tr{(\hat{\Pi}_f\hat{\Pi}_i)}},
\end{equation}
where $\hat{\Pi}_i=\ket{\psi_i}\bra{\psi_i}$ and $\hat{\Pi}_f=\ket{\psi_f}\bra{\psi_f}$ are the projectors on the initial (pre-selected) and final (post-selected) system states, respectively. The knowledge of the ancilla state conditioned on the successful post-selection of the final system state $\ket{\psi_f}$ provides information on the observable that is related to both the initial and final state, through the weak value. Weak values are unbounded complex numbers. In typical settings, experimental observations are proportional to the real or to the imaginary part of the weak value \cite{josza2007PRArealimWV}. The ancilla wavefunction shift may become very large when the pre- and post-selected states are nearly orthogonal, as can be seen from the denominator of equation \ref{eq:weak_value_definition}, a feature exploited extensively in the literature, called ``weak value amplification" \cite{aharonov1988result, dressel2014RMPweakvalue}. In practice, due to the weak system-ancilla coupling, the experimental determination of the weak value requires an average over many measurements, especially with large amplification, as the post-selection probability is approximately given by the overlap between the initial and final states ($\vert\langle\psi_f\ket{\psi_i}\vert^2$). 

The consideration of weak values in quantum measurement theory gave rise to many applications. Metrology \cite{zhang2015precision, Lundeen2020PRLperformanceWM, xukedemvaidmanPRL2013phaseim} and sensing \cite{steinberg2017natphysWMNLO, Qiu2016OLchiralWM, li2016AOglucoseWM, li2018chiral} benefit largely from amplification \cite{Lundeen2017PRLperformanceWM, Jordan2014PRXWVamp}. Harnessing weak value amplification evidenced new physical phenomena, such as minute optical effects in beam propagation \cite{hosten2008observation, jayaswal2013weak, gottedennis2011NJPbeamshiftsWM, ling2017RPPspinhalllight}, but also improved the control tiny experimental perturbations \cite{dixon2009PRLbeamdeflectionWM, boyd2014PRLsmallrotation} for sensitive experiments. In addition, weak values are useful to perform quantum tomography. Indeed, as complex numbers, they give a direct access to the complex components of the quantum state \cite{lundeenbamber2011nature, boyd2014natcommtomographyOAM, wu2013scireptomography}. Resulting from weakly perturbing measurements, weak values also provide essential insights into issues related to quantum foundations \cite{goggin2011PNASLeggetGarg, steinberg2012PRLheisenbergdist, matzkin2019weak}, such as paradoxes \cite{steinberg2004physletta3box, Yokota2009NJPHardy, denkmayr2014observation, jwpan2019PNASpigeon} or sensing quantum particles along trajectories \cite{kocsis2011observing, bliokh2013NJPtrajectoryPoynting, matzkin2012PRLtrajectory}. Weak values have also the potential to benefit quantum computing and quantum information processing \cite{lund2011efficient, martinezrincon2017QIPnlWM, pati2014annphysWMdiscord, gross2018qubit, weberdresseljordan2014natureWM, toddbrun2008PRAwmqubits}.

Although their experimental and practical usefulness is undeniable, there have been many difficulties and debates in interpreting weak values since their definition \cite{svensson2013pedagogical, dressel2014RMPweakvalue, matzkin2019weak}. In view of their significance, we provide a geometric description of weak values in order to clarify their meaning in dispassionate terms. As a direct reflection of their connection to experimental observations, weak values are usually discussed in terms of their real and imaginary parts \cite{svensson2013pedagogical, dressel2014RMPweakvalue, josza2007PRArealimWV}. Instead, we studied them in the polar representation, in terms of modulus and argument. This approach leads to a geometrical interpretation of the argument of weak values in terms of geometric phases. When a quantum system evolves, it gives rise to two different types of phases: the dynamical phase, related to the time evolution of the system, and the geometric phase (also called Pancharatnam phase, Berry phase or Pancharatnam-Berry phase), related to the intrinsic geometry of the system state space. It was discovered independently on several occasions \cite{kato1950adiabatic, pancharatnam1956generalized, longuet1958studies, berry1984quantal}. In the 50s, Pancharatnam defined the geometric phase as the difference of phase acquired by light along a cyclic polarization trajectory \cite{pancharatnam1956generalized}, which generalizes to quantum systems following an adiabatic and cyclic evolution, as Berry showed \cite{berry1984quantal}. Later on, it was discovered that neither the adiabatic condition nor the cyclic one were necessary {\cite{shapere1989geometric, mukundasimon1993AOPgmphase}. Any quantum system evolving from an initial state to a final one describes a geometrical phase \cite{cohen2019geometric}. Sj{\"o}qvist recognized the connection existing between weak values and geometric phases \cite{sjoqvist2006PLAgmphaseWM}. Tamate \emph{et al.} studied the geometric phase in quantum erasers and therefrom identified the Pancharatnam phase in weak measurements \cite{tamate2009NJPqueraserWM}.  Using the Majorana representation (which represents states of $N$-level system as $N-1$ stars on the Bloch sphere), Cormann and Caudano investigated the appearance of geometric phases in weak measurements for some specific cases, including Pauli matrices\footnote{Technically, they studied the case of modular values \cite{kedem2010PRLmodularvalue}, which appear in quantum controlled gates \cite{toddbrun2008PRAwmqubits}, where the weakly measured operator in (\ref{eq:weak_value_definition}) is unitary rather than Hermitian.} and weak values of projectors on pure states \cite{cormann2017geometric}. Ho and Imoto related the polar decomposition of modular values to geometric phases as well \cite{lebinho2018JMPHYSmodvalgmphase}. The appearance of geometric phases for weak measurements was also considered in the context of the spin Hall effect of light \cite{Samlan2017JOPTgmphasewm, Pal2019PRAgmphasewm} and for sequential measurements \cite{cho2019NatPhysGMphaseWM}. The link between weak values and the Bargmann invariant \cite{bargmann1964JMathPhysinvarriant} was noted on multiple occasions \cite{sjoqvist2006PLAgmphaseWM, tamate2009NJPqueraserWM, cormann2017geometric, gedik2021JPhysAgmphaseWM}. However, there is a lack of a general formalism to study the geometric phase of arbitrary weak values and to understand their geometrical properties in larger dimensions, beyond the special case of the qubit Bloch sphere.

In this paper, we provide for the first time to our knowledge a geometric description of the weak values of general observables in $N$-dimensional quantum systems. We especially describe how the argument of the weak value of an observable defines a geometric phase. The space of traceless Hermitian operators acting on an $N$-dimensional Hilbert space is of dimensions $N^2-1$. As a result, to any observable, we can associate a real vector on $S^{N^2-2}$, the unit sphere in $N^2-1$ dimensions. Since any pure state can be represented by a projector, a Hermitian operator, the pre-selected and post-selected states can also be assigned to points on  $S^{N^2-2}$. Thus, we describe any weak value starting from three real vectors on $S^{N^2-1}$. When the observable corresponds to a projector on a pure state, all three vectors belong in addition to the complex projective space $\mathbb{C}\textrm{P}^{N-1}$ \cite{bengtsson2017geometry}. This space has a representation as a non-trivial subspace of $S^{N^2-2}$, which generalizes the Bloch sphere \cite{kimura2003PLAblochnlevel, Bertlmann2008JPhysAblochqudits, goyal2016JPhysABlochQudits}. In the case of two-level systems,  i.~e. $\mathbb{C}\textrm{P}^{1}$,  the space of pure states is fully equivalent to the complete surface of the 2-sphere $S^2$: this is the well-known Bloch sphere representation of qubits. When considering three-level and higher level systems, their pure state space, $\mathbb{C}\textrm{P}^{2}$\dots, $\mathbb{C}\textrm{P}^{N-1}$, is not equivalent to the complete surface of a larger unit sphere. All pure states are on the surface of the sphere $S^{N^2-2}$ but all points on the surface do not represent a quantum state. The symplectic area\footnote{The symplectic area is defined in an even manifold that is formed by pairs of directions: for example, symplectic areas in classical mechanics are specified in terms of position and momentum \cite{siegel2014symplectic}. Here, we note that  $2N-2$ free parameters describe the states in an $N$-dimensional complex Hilbert space (considering the global phase and the normalization), which accounts for the even number of dimensions.} in $\mathbb{C}\textrm{P}^{N-1}$ created by the three vectors characterizing the pre-selected state, the probed projector observable, and the post-selected state is associated to the argument of the weak value. Indeed, for three projectors on pure states of an $N$-level system, the argument of the Bargmann invariant $\Tr(\hat{\Pi}_3\hat{\Pi}_2\hat{\Pi}_1)$ defines a geometric phase associated to the symplectic area on the generalized Bloch-sphere $(\mathbb{C}\textrm{P}^{N-1})$ \cite{hangan1994GeomDBargmannSymplectic, ortega2003JPhysAtrigoCP2, mukunda2003bargmann}. This area is wholly specified by the geodesic triangle spanned by the three vectors representing these three states. Besides, the argument of the Bargmann invariant of three pure states coincides with the argument of the weak value of the projector $\hat{\Pi}_2$, when the pre-selected and post-selected states are given by the projectors $\hat{\Pi}_1$ and $\hat{\Pi}_3$, respectively (see third member of (\ref{eq:weak_value_definition})). Thus, the symplectic area charaterizes the argument of all weak values of projectors. In a further step, we show that the argument of the weak value of any observable is equivalent to the argument of an effective Bargmann invariant, arising from the weak value of a very specific projector. Therefore, we obtain that the argument corresponds to a geometric phase. However, the latter is then expressed starting from three real vectors on $S^{N^2-2}$, of which only two are still constrained to the $\mathbb{C}\textrm{P}^{N-1}$ submanifold. 

The following of the paper is organized as follows. In order to illustrate the concepts introduced and to establish connections to the existing literature on the use of generalized Bloch spheres, we first derive the expression of weak values of projectors on pure states in qutrit systems in terms of vectors on $S^{7}$ that also belong to $\mathbb{C}\textrm{P}^{2}$ and we find the weak value argument. In a second step, we generalize these results to projectors in $N$-level systems. We then determine the expression of weak values of general observables in terms of real vectors on $S^{N^2-2}$. We also show the expression of their argument in terms of the argument of an effective Bargmann invariant involving a particular projector. Then, we consider the particular case of the generators of SU($N$) given by the generalized Gell-Mann matrices, which are traceless Hermitian observables generalizing the Pauli matrices in $N$-level systems. Finally, before concluding, we consider a few situations to apply our formalism to: we study the case of a general observable in SU(2); we discuss projectors on degenerate subspaces and Hermitian quantum gates; and we illustrate the expressions for average values and the quantum uncertainties that result from considering identical pre- and post-selected states. In the end, a few technical calculations were put in annexes, along with auxiliary contextualization of our work. In particular, in \ref{appendix_A} and \ref{appendix_B}, we explicit our conventions used for SU($N$) generators, projectors, and the star product. We consigned the details of the computation of the weak value to \ref{appendix_C}. Finally, in \ref{appendix_D} and \ref{appendix_E}, we provide additional information about the geometry of states on the generalized Bloch spheres and about the related star and wedge products.

\section{Weak values of projectors in $\mathbb{C}^3$}\label{section:projC3}

In this section we focus on weak values $\Pi_{r,w}$ of three-level system projectors on pure states. We describe them in terms of the eight $3\times3$ Gell-Mann matrices $\hat\lambda_i$, a representation of the Lie algebra $\textrm{SU}(3)$ \cite{macfarlane1968gell} (we define the generators in \ref{appendix_A}). Pure states of two-level quantum systems belong to the complex projective line $\mathbb{C}\textrm{P}^1$ and can be represented on the surface of the Bloch sphere (2-sphere with a unit radius). In the case of three-level systems, pure states are associated with rays that are points of the projective plane $\mathbb{C}\textrm{P}^2$. This space is not equivalent to the surface of the 7-sphere: all three-level states are on the sphere surface but most points on the surface of the 7-sphere are not proper quantum states. A projector on a pure state is 
\begin{equation}\label{eq:proj3def}
\hat{\Pi}_r=\frac{1}{3}\left(\hat{I}+\sqrt{3}\,\vec{r}\cdot\hat{\vec{\lambda}}\right),
\end{equation} 
where $\hat{\vec{\lambda}}$ is a vector whose components are the eight Gell-Mann matrices and $\vec{r}$ is an eight-dimensional, normalized\footnote{In this work, we chose the convention to work with normalized vectors on a single unit sphere. See \ref{appendix_B} for more details.}, real vector \cite{mallesh1997generalized, khanna1997AnnPhysGMphaseSU3}. Additionally, the vector $\vec{r}$ must verify  $\vec{r}\star\vec{r}=\vec{r}$, where the symmetric star product is defined by $(\vec{\alpha}\star\vec{\beta}\,)_c=\sqrt{3}d_{abc} \alpha_a \beta_b$ \cite{mallesh1997generalized, khanna1997AnnPhysGMphaseSU3} (the components $a,b,c \in \{1,2\dots 8\}$ and Einstein's summation convention is assumed for repeated indices). The constants $d_{abc}$ are completely symmetric under index permutation and arise from the anti-commutator of the generators of the of SU(3) Lie algebra: they are determined by $d_{abc}=\frac{1}{4}\Tr(\hat\lambda_a\{\hat\lambda_b,\hat\lambda_c\})$. The star product condition ensures that the vector $\vec{r}$ on the unit 7-sphere also corresponds to a point of $\mathbb{C}\textrm{P}^2$ (hence represents a pure quantum state). This condition comes by imposing that the square of the projector should be equal to the projector itself (see \ref{appendix_B}). The star product appears in few papers in the literature  \cite{goyal2016JPhysABlochQudits, mallesh1997generalized, khanna1997AnnPhysGMphaseSU3, byrd1998JMPHYSsu3starp, Siennicki2001AnnPhysStarWedge, Byrd2003PRAstardimN, graf2021PRBstarpcrossp} but not many (to our knowledge). We would like to stress that many familiar properties of the SU(2) Bloch sphere do not hold in larger dimensions. For example, three orthogonal basis states of a three-level system are associated to three 8-dimensional vectors forming angles of $120^{\circ}$ in a single plane (on the Bloch sphere, two vectors associated to orthogonal states are opposite, that is form $180^{\circ}$ angles). Also, there is no equivalent of the $\star$ product in SU(2). For this reason, we discuss a few essential properties of the state representation as generalized Bloch spheres in \ref{appendix_D} and \ref{appendix_E} to provide more context for the interested reader. 

We use the properties of the generators of the Lie algebra of $\textrm{SU}(3)$ to calculate the expression of the weak value of a three-level system projector from (\ref{eq:weak_value_definition}):  $\Pi_{r,w}=\Tr\, (\hat{\Pi}_f\hat{\Pi}_r\hat{\Pi}_i)\, / \Tr\, (\hat{\Pi}_f\hat{\Pi}_i)$. Applying the definition of projectors in terms of Gell-Mann matrices (\ref{eq:proj3def}), we find
\begin{equation}
        \label{eq:weak_value_projector_Su3}
        \Pi_{r,w}=\frac{1+2\, \vec{f}\cdot\vec{r}+2\, \vec{r}\cdot\vec{i}+2\, \vec{f}\cdot\vec{i}+2\,\vec{f}\cdot\left(\vec{r}\star\vec{i}\right)+\textrm{i}\,2\sqrt{3}\,\vec{f}\cdot\left(\vec{r}\wedge\vec{i} \right)}{3+6\,\vec{f}\cdot\vec{i}},
\end{equation}
where the 8-dimensional real vectors $\vec{i}$, $\vec{f}$ and $\vec{r}$ represent the pre- and post-selected states and the weakly measured projector state, respectively. In addition to the previously defined star product, we also introduced the antisymmetric wedge product \cite{mallesh1997generalized, Siennicki2001AnnPhysStarWedge, graf2021PRBstarpcrossp} $(\vec{\alpha}\wedge\vec{\beta})_c=f_{abc} \alpha_a \beta_b$, with $f_{abc}=-\frac{1}{4} \textrm{i}\Tr(\hat\lambda_a [\hat\lambda_b, \hat\lambda_c ])$, the structure constants of the Lie algebra of $\textrm{SU}(3)$ (which are completely anti-symmetric under index permutations and emerge from the commutator of the generators of the Lie algebra). Note that the $\star$ and $\wedge$ products produce vectors that are both outside of $\mathbb{C}\textrm{P}^2$ and not normalized in general (so they do not represent states). The vector $\vec{\alpha}\wedge\vec{\beta}$ produced by the wedge product is orthogonal to the two initial vectors $\vec{\alpha}$ and $\vec{\beta}$. We also have that the two products are orthogonal: for any pre- and postselected states $(\vec{i}\wedge\vec{f})\cdot (\vec{i}\star\vec{f})=0$. In (\ref{eq:weak_value_projector_Su3}), the two scalar products $\vec{f}\cdot(\vec{r}\star\vec{i})$ and $\vec{f}\cdot(\vec{r}\wedge\vec{i})$ are invariants under cyclic permutations of the three vectors and under unitary transformations. The qutrit weak value (\ref{eq:weak_value_projector_Su3}) bears similarities with the qubit case \cite{cormann2017geometric}: for SU(2), the structure constants $f_{abc}$ are given by the Levi-Cevita symbol while $d_{abc}=0$, so that the wedge product reduces to the usual cross-product in three dimensions while the star product contribution disappears. We discuss additional properties of the $\star$ and $\wedge$ products in \ref{appendix_D} and we provide detailed calculations leading to (\ref{eq:weak_value_projector_Su3}) in \ref{appendix_C} (although we recommend skipping the latter for now).

Weak values are regularly considered in terms of their real and imaginary parts because this is how they affect typical weak measurements. Nonetheless, interpreting weak values in terms of their modulus and argument provides us more insight about their geometrical properties. Considering the real and imaginary parts of (\ref{eq:weak_value_projector_Su3}), the weak value argument is,  
\begin{eqnarray}
        \label{eq:argument_weak_value_projector_Su3}
       \arg \Pi_{r,w}=
       \arctan\frac{2\sqrt{3}\,\vec{f}\cdot\left(\vec{r}\wedge\vec{i} \right)}{1+2\left(\vec{f}\cdot\vec{r}+\vec{r}\cdot\vec{i}+\vec{f}\cdot\vec{i}\right)+2\,\vec{f}\cdot\left(\vec{r}\star\vec{i} \right)}+\phi\left(\Pi_{r,w}\right),
\end{eqnarray}
where the term $\phi$ essentially determines the appropriate quadrant (which depends on the signs of the real and imaginary parts of the weak value):
\begin{equation}
\label{eq:expression_phi}
\phi\left(f\right)=\left\lbrace\begin{array}{c} 0~\textrm{if}~\Re\left(f\right)>0 \\ \pi~\textrm{if}~\Re\left(f\right)<0
\end{array}\right. .
\end{equation}
This argument corresponds to a geometric phase because it is equal to the argument of the Bargmann invariant \cite{bargmann1964JMathPhysinvarriant} associated to  the initial state, the projector state weakly measured and the final state \cite{sjoqvist2006PLAgmphaseWM, tamate2009NJPqueraserWM, cormann2017geometric, gedik2021JPhysAgmphaseWM}. Indeed, in the weak value expression $\Pi_{r,w}=\Tr\, (\hat{\Pi}_f\hat{\Pi}_r\hat{\Pi}_i)\, / \Tr\, (\hat{\Pi}_f\hat{\Pi}_i)$, the numerator is equal to the Bargmann invariant while the denominator $\Tr\, (\hat{\Pi}_f\hat{\Pi}_i)=|\bra{\psi_f}\psi_i\rangle|^2$ is always positive. The argument of the Bargmann invariant describes a geometric phase equal to half the symplectic area of the geodesic triangle with the vertices $\vec{f}, \vec{r}$ and $\vec{i}$ in the projective space $\mathbb{C}\textrm{P}^2=\textrm{SU}(3)/\textrm{U}(2)$ \cite{mallesh1997generalized}. Hence, the argument of the weak value of a projector in three-level systems corresponds as well to half the symplectic area of a geodesic triangle in the $\mathbb{C}\textrm{P}^2$ manifold \cite{hangan1994GeomDBargmannSymplectic, ortega2003JPhysAtrigoCP2, mukunda2003bargmann}. These geometric insights generalize the results obtained for weak values of projectors in qubit systems: in $\mathbb{C}\textrm{P}^1$, the argument is related to the solid angle subtended by the three vectors on the Bloch sphere: $\arg \Pi_{r,w}=-\frac{1}{2}\Omega_{irf}$ \cite{sjoqvist2006PLAgmphaseWM, tamate2009NJPqueraserWM, cormann2017geometric}. The complex projective spaces $\mathbb{C}\textrm{P}^n$ are Kähler manifolds  \cite{bengtsson2017geometry}, which means that they are equipped with a Hermitian form that provides both a Riemannian metric (Fubini-Study) and a symplectic form. In the qubit case, the Riemannian and symplectic areas of geodesic triangles are identical. However, this is no longer the case in $\mathbb{C}\textrm{P}^2$. For this reason, in three-level systems, the weak value argument cannot be as straightforwardly interpreted as a solid angle, independently of one's preferred geometrical representation of $\mathbb{C}\textrm{P}^2$. Nevertheless, the symplectic area is independent of the particular surface on which it is computed. The argument of the weak value is thus appropriately seen as arising from a contour integral along the boundary geodesics of the triangle, which accumulates the incremental geometric phase changes along the closed path. The argument of the projector weak value is a three-point invariant of the geodesic triangle.

\begin{figure} [tb]
\centering 
\includegraphics[width=1\textwidth]{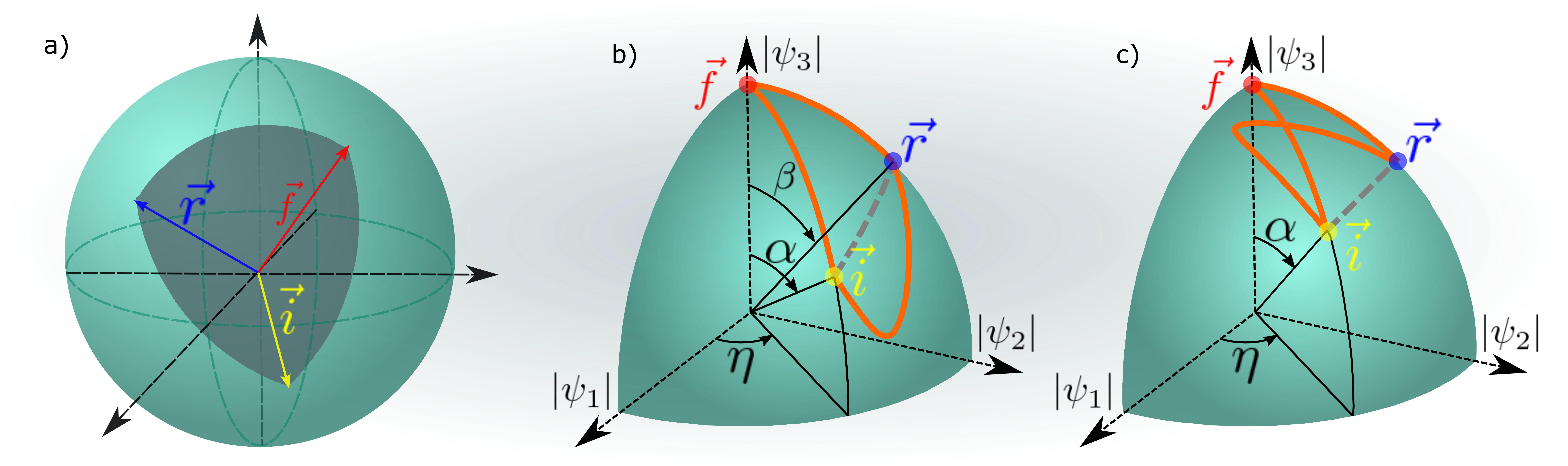}
  \caption{Representation of the geodesic triangle generated by the initial state ($\vec{i}$), the observable state ($\vec{r}$) and the final state ($\vec{f}$) in the complex projective spaces $\mathbb{C}\textrm{P}^1$ (a) and $\mathbb{C}\textrm{P}^2$ (b and c). The argument of the weak value of a projector is associated to the symplectic area of the triangle. (a) For two-level systems, the argument is also linked to the area and solid angle of the spherical triangle spanned by the three $\mathbb{R}^3$ vectors representing states on the Bloch sphere. (b) and (c) Depiction of two different geodesic triangles of three-level systems, using the spherical octant projection.\label{fig:ray_space_triangle}}
\end{figure}

In figure \ref{fig:ray_space_triangle}, we represent the geodesic triangle spanned by three quantum states associated to a projector weak value, in $\mathbb{C}\textrm{P}^1$ for qubit systems and in $\mathbb{C}\textrm{P}^2$ for qutrit systems. For two-level systems, the three quantum states lay on the surface of the Bloch sphere and the argument is connected to the area of the spherical triangle, equivalent to the solid angle. For three-level systems, the geometry is more complicated. Most generally, we can represent $\mathbb{C}\textrm{P}^2$ graphically using a three-dimensional sphere octant, where each point is furthermore associated to a torus linked to the two phase components $\chi_2$ and $\chi_3$ of the state vector in Hilbert space \cite{bengtsson2017geometry, mallesh1997generalized}: $\ket{\psi}=\left(|\psi_1| e^{\mathrm{i}\chi_1}, |\psi_2| e^{\mathrm{i}\chi_2}, |\psi_3| \right)^T$. Each state $\ket{\psi}$ is projected on the sphere octant on the point with (real) coordinates $\vec{q}=\left(|\psi_1|, |\psi_2|, |\psi_3|\right)^T$. Considering the invariance under unitary transformations, the most general geodesic triangle can be represented by the states $\ket{f}=\left(0,0,1\right)^T$, $\ket{r}=\left(0,\sin{\beta}, \cos{\beta}\right)^T$ and $\ket{i}=\left(\sin{\alpha}\cos{\eta},e^{i\chi_2}\sin{\alpha}\sin{\eta}, \cos{\alpha} \right)^T$, as depicted in figure \ref{fig:ray_space_triangle}. Beware that the geodesics connecting the triangle vertices on the octant do not typically appear as spherical arcs. Each pair of vertices generates a unique complex projective line (i.e. a two dimensional subspace isomorphic to a Bloch sphere) in the complex projective plane $\mathbb{C}\textrm{P}^2$. Topologically, the three geodesics connecting the vertices are thus arcs of great circles in each of these three distinct (in general) $\mathbb{C}\textrm{P}^1$ subspaces. However, when projected on the sphere octant, the geodesic triangle is inevitably distorted. On the $S^7$ sphere, these geodesics would appear as circle arcs connecting the vertices. However these circle arcs would not be arcs of great circles because the geodesics of $\mathbb{C}\textrm{P}^2$ are not those of $S^7$ (which is a reminder that the points of $\mathbb{C}\textrm{P}^2$ are constrained to a four-dimensional subspace of $S^7$ equipped with the Fubini-Study metric and not with the metric of the round sphere). The general expression of a state on $S^7$ and of the geodesic arc linking $\vec{r}$ and $\vec{f}$ are given in \ref{appendix_E} for reference.

\section{Weak values of projectors in $\mathbb{C}^N$}\label{section:projCN}

Our results on three-level systems can be easily generalized to $N$-level systems. Projectors on pure states in $\mathbb{C}^N$ are associated with straight complex lines passing through the origin or, equivalently, with rays that are points of the projective space $\mathbb{C}\textrm{P}^{N-1}$. We describe them in terms of the $N^2-1$ generators of the Lie algebra of $\textrm{SU}(N)$ that generalize the Pauli and Gell-Mann matrices (see \ref{appendix_A} for details). In that case, a projector on a pure state is (see \ref{appendix_B})
\begin{equation}\label{projNdef}
\hat{\Pi}_r=\frac{1}{N}\hat{I}_N+\sqrt{\frac{N-1}{2N}}\,\vec{r}\cdot\hat{\vec{L}}, 
\end{equation}
where $\hat{\vec{L}}$ is a vector whose $N^2-1$ components give the generators $\hat{L}_a$. The real vector $\vec{r}$ has also $N^2-1$ components. It is normalized ($\vec{r}\cdot\vec{r}=1$), so that the state is on the surface of the $S^{N^2-2}$ sphere in $N^2-1$ dimensions. Additionally it is constrained to a subspace of this sphere by the extra condition $\vec{r}\star\vec{r}=\vec{r}$, where the star product is defined by $(\vec{q}\star\vec{r})_c=\sqrt{\frac{N\left(N-1\right)}{2}}\frac{1}{N-2}d_{abc}q_a r_b$ \cite{Byrd2003PRAstardimN}, with the symmetric constants of SU(N) given from the anti-commutator by $d_{abc}=\frac{1}{4}\Tr (\hat{L}_a\{\hat{L}_b,\hat{L}_c\})$ as previously. This constraint ensures that the point $\vec{r}$ on the sphere corresponds to a projector on a pure state and, thus, is associated to a proper quantum pure state belonging to $\mathbb{C}\textrm{P}^{N-1}$. As for  qutrit systems, this condition makes it impossible to describe the geometry of the states as the complete surface of a sphere: the pure quantum states belong to a $(2N-2)$-dimensional subspace of the $S^{N^2-2}$ sphere. In this space, when two quantum states $\ket\psi$ and $\ket\phi$ are orthogonal, the vectors $\vec{r}_\psi$ and $\vec{r}_\phi$  representing those states on the generalized Bloch sphere form an angle of $\pi-\arccos{\frac{1}{N-1}}$. In the limit of very large N, the angle between the vectors associated to orthogonal states tends to $90^\circ$. In \ref{appendix_B}, we provide context to our definition of the symmetric $\star$ product in SU($N$) as a generalization of its original SU(3) definition \cite{mallesh1997generalized, khanna1997AnnPhysGMphaseSU3}.

The argument of the weak value of a projector $\hat{\Pi}_r$ on a pure state of an $N$-level system is expressed using the properties of the generators of $\textrm{SU}(N)$ as,
\begin{eqnarray}
     \label{eq:argument_weak_value_projectors_Sun}
   \arg \Pi_{r,w}&=&
    \arctan{\frac{2\left(\frac{N-1}{2N}\right)^{\frac{3}{2}}\vec{f}\cdot\left(\vec{r}\wedge\vec{i}\right)}{\frac{1}{N^2}+\frac{N-1}{N^2}\left(\vec{f}\cdot\vec{r}+\vec{r}\cdot\vec{i}+\vec{f}\cdot\vec{i}\right)+\frac{(N-1)(N-2)}{N^2}\vec{f}\cdot\left(\vec{r}\star\vec{i}\right)}} \nonumber\\ 
     &+& \phi\left(\Pi_{r,w}\right), 
\end{eqnarray}
where $\phi\left(\Pi_{r,w}\right)$ selects the appropriate quadrant and is specified in (\ref{eq:expression_phi}). The anti-symmetric wedge product is defined similarly to the qutrit case using the structure constants of SU($N$), which are obtained from the commutation relationships of the generators:  $f_{abc}=-\frac{1}{4}\textrm{i}\Tr(\hat{L}_a [\hat{L}_b,\hat{L}_c])$. It produces the vector with components $(\vec{\alpha}\wedge\vec{\beta})_c=f_{abc} \alpha_a \beta_b$, which is orthogonal to both $\vec{\alpha}$ and $\vec{\beta}$. The expression (\ref{eq:argument_weak_value_projectors_Sun}) of the argument of the weak value generalizes the one obtained for three-level systems (\ref{eq:argument_weak_value_projector_Su3}) to the case of $N$-level system projectors on pure states. Taking $N=3$ and $N=2$, we recover the results for three-level system projectors and qubit systems, respectively. As explained previously, the argument of the weak value of a projector is equal to the argument of the Bargmann invariant of the three states involved. Therefore, the argument (\ref{eq:argument_weak_value_projectors_Sun}) represents a geometric phase: it is essentially the expression of half the symplectic area of the geodesic triangle in $\mathbb{C}\textrm{P}^{N-1}$. Using an appropriate unitary transformation, any group of three states of $\mathbb{C}\textrm{P}^{N-1}$ can be mapped to a  $\mathbb{C}\textrm{P}^{2}$ subspace. Therefore, all observations valid for the weak value of qutrit projectors on pure states extend to arbitrary larger dimensions. In particular, figure \ref{fig:ray_space_triangle} provides a valid representation of geodesic triangles in $\mathbb{C}\textrm{P}^{N-1}$. We note that there are two three-point invariants that contribute to the geometric phase:  $\vec{f}\cdot(\vec{r}\wedge\vec{i})=f_{abc}f_a r_b i_c$ and $\vec{f}\cdot(\vec{r}\star\vec{i})=\sqrt{\frac{N\left(N-1\right)}{2}}\frac{1}{N-2} d_{abc}f_a r_b i_c$. Both are invariant under cyclic permutations of the three vectors and under unitary transformations. However, the former is anti-symmetric under the permutation of two vectors, while the latter is symmetric. Detailed calculations leading to (\ref{eq:argument_weak_value_projectors_Sun}) are provided in \ref{appendix_C}.

\section{Weak values of general observables}
Weak measurement are not confined to projectors on pure states. In practice, experiments also deal the weak values of arbitrary observables, such as the spin of a particle, projectors on degenerate subspaces or compound observables in multipartite systems to name a few. In this section, we study the weak value of a general Hermitian observable $\hat{A}$ of an $N$-level system. We will show that the the weak value of any observable can be expressed in terms of the weak value of a very specific projector that will provide us with a geometrical description in the spirit of what we did before. We decompose the observable in terms of the $N \times N$ identity operator and a traceless operator from $\textrm{SU}(N)$:
\begin{equation}\label{eq:generalObservableAdefinition}
\hat{A}=a_I\, \hat{I}_N+a_L\, \vec{\alpha}\cdot\hat{\vec{L}},
\end{equation}
where $\hat{\vec{L}}$ is a vector whose $N^2-1$ components are the generators of the $\textrm{SU}(N)$ Lie group (see \ref{appendix_A}), $a_I$ and $a_L$ are real constants and $\vec\alpha$ is a normalized vector with $N^2-1$ real components. Using the properties of the generators $\hat{\vec{L}}$ to compute the traces appearing in the definition (\ref{eq:weak_value_definition}), we obtain the weak value of a general observable (as shown in \ref{appendix_C}):
\begin{eqnarray}
    \label{eq:expression_weak_value_general_observable}\nonumber
   A_{\alpha,w}&=&\frac{1}{\frac{1}{N}+\frac{N-1}{N}\vec{f}\cdot\vec{i}} \Bigg[\textrm{i}\frac{a_L\left(N-1\right)}{N}\vec{f}\cdot\left(\vec{\alpha}\wedge\vec{i}\right) \\ \nonumber
   &+& \frac{a_I}{N}+\frac{a_I\left(N-1\right)}{N}\vec{f}\cdot\vec{i} +\frac{a_L\sqrt{2\left(N-1\right)}}{N\sqrt{N}}\left( \vec{f}\cdot\vec{\alpha}+\vec{\alpha}\cdot\vec{i}\right) \\
    &+&\frac{a_L\sqrt{2\left(N-1\right)}\left(N-2\right)}{N\sqrt{N}}\vec{f}\cdot\left(\vec{\alpha}\star\vec{i}\right)\Bigg].
\end{eqnarray}
This expression generalizes our previous results on projectors, which we recover by setting the appropriate coefficients $a_I$ and $a_L$ from (\ref{projNdef}) and by constraining the operator using $\vec\alpha=\vec\alpha\star\vec\alpha$. It is thus crucial to note that, in (\ref{eq:expression_weak_value_general_observable}), the vectors $\vec{i}$ and $\vec{f}$ must obey the constraints $\vec{i}=\vec{i}\star\vec{i}$ and $\vec{f}=\vec{f}\star\vec{f}$ because they represent states, while $\vec\alpha$ is allowed to range freely on the whole surface of the $S^{N^2-2}$ sphere (for this reason, we denote the latter vector with a greek letter, while the former are identified by roman letters). Our normalization conventions for the projector definition (\ref{projNdef}) allowed us to express the weak values in terms of the geometrical properties of three vectors that all belong to the same unit sphere (see \ref{appendix_B}).

By considering the ratio of the imaginary and real parts of (\ref{eq:expression_weak_value_general_observable}), we find the argument of the weak value of an arbitrary Hermitian observable:
\begin{eqnarray}
     \label{eq:argument_weak_value_general_observable}
    \textrm{arg} \left( A_{\alpha,w} \right)= \phi\left(A_{\alpha,w}\right) +\\
     \arctan \frac{\frac{a_L\left(N-1\right)}{N} \vec{f}\cdot\left(\vec{\alpha}\wedge\vec{i}\right) }{\frac{a_I}{N}+\frac{a_I\left(N-1\right)}{N}\vec{f}\cdot\vec{i}+\frac{a_L\sqrt{2\left(N-1\right)}}{N\sqrt{N}}\left(\vec{f}\cdot\vec{\alpha}+\vec{\alpha}\cdot\vec{i}\right)+\frac{a_L\sqrt{2\left(N-1\right)}\left(N-2\right)}{N\sqrt{N}}\vec{f}\cdot\left(\vec{\alpha}\star\vec{i}\right)}, \nonumber
\end{eqnarray}
where $\phi\left(A_{\alpha,w}\right)$ specifies the quadrant according to (\ref{eq:expression_phi}). This expression bears similarities to the argument of the weak value of a projector on a pure state (\ref{eq:argument_weak_value_projectors_Sun}). Notwithstanding the appearance of the constant factors $a_I$ and $a_L$, the essential differences are the new term $a_I/N$ in the denominator and the fact that $\vec\alpha$ does not represent a state in general. We notice that the numerator is fully antisymmetric under permutations of vectors, while the denominator is symmetric under permutations of the initial and final states. The term involving the star product is even fully symmetric under permutations of the three vectors. 

In order to interpret this argument in terms of a geometric phase, we relate it to the weak value of a particular projector on a pure state (we call it $\hat\Pi_{i'}$).This approach allows us to link the argument to the Bargmann invariant of the pre-selected state, the projector $\hat\Pi_{i'}$ and the post-selected state, and, therefore, to a symplectic area in $\mathbb{C}\textrm{P}^{N-1}$. We define this projector by 
\begin{equation}\label{eq:iprimeProjectorDefinition}
    \hat{\Pi}_{i'}=\frac{\hat{A}\ket{\psi_i}\bra{\psi_i}\hat{A}}{\bra{\psi_i}\hat{A}^2\ket{\psi_i}}
\end{equation}
when $\hat{A}\ket{\psi_i}\ne 0$ (else the weak value is 0 and the argument is undefined anyway). The state $\ket{\psi_{i'}}$ results from the application of the observable to the initial state. Since the argument of the weak value of a projector is equal to the argument of the Bargmann invariant, we have
\begin{equation}
\label{eq:argument_pi_ip}
    \arg \Pi_{i',w}=\arg [\Tr\,(\hat{\Pi}_{f}\hat{\Pi}_{i'}\hat{\Pi}_{i})]
    =\arg\frac{\bra{\psi_f}\hat{A}\ket{\psi_i}\bra{\psi_i}\hat{A}\ket{\psi_i}\langle\psi_i\ket{\psi_f}}{\bra{\psi_i}\hat{A}^2\ket{\psi_i}}.
\end{equation}
The expectation value of $\hat{A}^2$ in the pre-selected state is strictly positive and does not contribute to the total argument: $\arg \bra{\psi_i}\hat{A}^2\ket{\psi_i}=0$. The average value $\langle A\rangle_{\psi_i}=\bra{\psi_i}\hat{A}\ket{\psi_i}$ is a real number: its argument is 0 if it is positive and $\pi$ if it is negative. Hence, the argument of the weak value of the observable $\hat{A}$ is equivalent to the argument of the weak value of the projector $\hat{\Pi}_{i'}$ modulo $\pi$:
\begin{equation}\label{eq:weakvalueArgumentWithipreime}
    \arg{A_w}=\arg{\Pi_{i',w}}-\arg{\langle A\rangle_{\psi_i}}.
\end{equation}
We find two contributions. First, the geometric phase arising from the geodesic triangle in $\textrm{CP}^{N-1}$ whose vertices correspond to the vectors $\vec i$, $\vec i'$ and $\vec f$ on the $S^{N^2-2}$ sphere. It is equal to half the symplectic area of the geodesic triangle. Second, another geometric phase that is given by the sign of average value of the observable in the initial state. It can easily be expressed in the present formalism, on $S^{N^2-2}$, in terms of the vectors $\vec i$ and $\vec \alpha$ by setting $\vec i =\vec f$ in (\ref{eq:expression_weak_value_general_observable}), as done later in section \ref{section:uncertainties} with (\ref{eq:averageValueOnSphere}).

Describing geometrically the relationship existing between an arbitrary initial state $\ket{\psi_i}$ and the quantum state associated to the projector $\hat\Pi_{i'}$ produced by an observable $\hat{A}$ according to (\ref{eq:iprimeProjectorDefinition}) is far from a trivial task in general. In the case of the Pauli observables $\vec{\alpha}\cdot\hat{\vec{\sigma}}$, the vector $\vec{i}'$ corresponds to the mirror image of the initial state $\vec{i}$ with respect to the axis $\vec\alpha$ of the Pauli operator (i. e. the direction of the spin measurement). However, these particular observables are also unitary operators, which helps in determining their action on a general initial state. More complicated operators or families of operators should be studied on case by case basis. Subsequently in this paper, we will consider in depth the situation of general observables of two-level systems and leave most higher dimensional cases for follow-up studies.

\section{Weak values of the generators of the Lie group of $\textrm{SU}(N)$}
As a relevant, direct application of the results obtained in the previous section, we express the weak values of the generators of the Lie algebra of $\textrm{SU}(2)$ (Pauli matrices), $\textrm{SU}(3)$ (Gell-Mann matrices) and $\textrm{SU}(N)$ in general. Operators linked to these generators describe the spin of particles, light polarization, the orbital angular momentum of light and the polarization correlations in entangled photons, to cite the most typical laboratory use. They are also related to observables in particle physics and cosmology, where conceptual applications of weak measurements start to emerge \cite{portosilva2021EPJCneutrinoWV}. We obtain the weak value and its argument, of all the generators, by setting $a_I=0$ and $a_L=1$ in the expressions (\ref{eq:expression_weak_value_general_observable}) and (\ref{eq:argument_weak_value_general_observable}), respectively.  

First, we recover the weak value and its argument for two-level systems ($N=2$) \cite{cormann2016revealing}, which are useful points of comparison: 
\begin{eqnarray}
      \label{eq:weak_value_pauli}
        \sigma_{r,w}&=&\frac{\vec{f}\cdot\vec{r}+\vec{r}\cdot\vec{i}+\textrm{i}\left[\vec{f}\cdot\left(\vec{r}\times\vec{i} \right)\right]}{1+\vec{f}\cdot\vec{i}},\\ 
        \arg \sigma_{r,w}&=&
        \label{eq:argument_weak_value_pauli}\arctan \frac{ \vec{f}\cdot\left(\vec{r}\times \vec{i}\right)}{\vec{f}\cdot\vec{r}+\vec{r}\cdot\vec{i}}
        +\phi\left(\sigma_{r,w}\right),
\end{eqnarray}
where $\phi\left(\sigma_{r,w}\right)$ is defined in (\ref{eq:expression_phi}) and the generator is given by $\vec{r}\cdot\hat{\vec{\sigma}}$. In the particular case of SU(2), the star product is null and wedge product is equivalent to the cross product because the structure constants $f_{abc}$ become the Levi-Civita symbol. Additionally, the vector $\vec{r}$ also represents a state and it is then possible to show that (\ref{eq:argument_weak_value_pauli}) is given by the sum of two solid angles connected to two Bargmann invariants \cite{cormann2016revealing}. These two solid angles correspond to the two contributions found in (\ref{eq:weakvalueArgumentWithipreime}) when using the projector $\hat\Pi_{i'}$ to determine the geometric phase. 

Second, we consider the generators $\vec\alpha\cdot\hat{\vec\lambda}$ of the Lie group $\textrm{SU}(3)$, i. e. the Gell-Mann matrices \cite{macfarlane1968gell}:
\begin{eqnarray}
    \label{eq:gellmann_weak_value}
    \lambda_{\alpha,w}&=& \frac{2}{\sqrt{3}}\frac{\vec{f}\cdot\vec{\alpha}+\vec{\alpha}\cdot\vec{i}+\vec{f}\cdot\left(\vec{\alpha}\star\vec{i}\right)+\sqrt{3}\, \textrm{i}\,\vec{f}\cdot\left(\vec{\alpha} \wedge \vec{i}\right)}{1+2\vec{f}\cdot\vec{i}},
\\
    \label{eq:expression_argument_gell_mann}
  \arg \lambda_{\alpha,w} &=& \arctan \frac{\sqrt{3}\, \vec{f}\cdot\left(\vec{\alpha}\wedge\vec{i}  \right)   }{\vec{f}\cdot\vec{\alpha}+\vec{\alpha}\cdot\vec{i}+\vec{f}\cdot\left(\vec{\alpha}\star\vec{i}\right)}+\phi\left(\lambda_{\alpha,w}\right),
\end{eqnarray}
with $\phi\left(\lambda_{\alpha,w}\right)$ defined in (\ref{eq:expression_phi}) and where $\vec\alpha$ does not represent a state in general. We note the elegant similarities between the weak values of qubit and qutrit systems. However, the complexity introduced in particular by the additional term involving the star product makes it no longer possible to interpret straightforwardly the geometric phase (\ref{eq:expression_argument_gell_mann}). Indeed, in addition to the three initial vectors $\vec{i}$, $\vec\alpha$ and $\vec{f}$, we also have to consider the directions of $\vec{f}\wedge\vec \alpha$ and $\vec{f}\star\vec\alpha$, which prevent us from reducing the problem to the three dimensions spanned by $\vec{i}$, $\vec\alpha$ and $\vec{f}$.

Third, for the generators $\vec\alpha\cdot\hat{\vec L}$ of $\textrm{SU}(N)$ generalizing the Pauli and Gell-Mann matrices, the weak value and its argument are
\begin{equation}
    L_{\alpha,w}=\sqrt{2\frac{N-1}{N}} \frac{\vec{f}\cdot\vec{\alpha}+\vec{\alpha}\cdot\vec{i}+(N-2)\vec{f}\cdot\left(\vec{\alpha}\star\vec{i}\right)+\textrm{i} \sqrt{\frac{N^2-N}{2}}\vec{f}\cdot\left(\vec{\alpha}\wedge\vec{i}\right)}{1+(N-1) \vec{f}\cdot\vec{i}},
    \label{eq:expression_weak_value_generators_suN}
\end{equation}
\begin{equation}
     \arg L_{\alpha,w} = \arctan \frac{\sqrt{\frac{1}{2}N(N-1)}\ \vec{f}\cdot\left(\vec{\alpha}\wedge\vec{i}\right) }{\vec{f}\cdot\vec{\alpha}+\vec{\alpha}\cdot\vec{i}+(N-2)\ \vec{f}\cdot\left(\vec{\alpha}\star\vec{i}\right)}+\phi\left(L_{\alpha,w}\right),
     \label{eq:argument_weak_value_generators_suN}
\end{equation}{}
with $\phi\left(L_{\alpha,w}\right)$ defined in (\ref{eq:expression_phi}). 

\section{Weak value of a two-level system observable}
We turn our attention to the weak values of arbitrary observables in two-level systems. In particular, we are interested in analyzing their connection to the projector $\hat\Pi_{i'}$ (\ref{eq:iprimeProjectorDefinition}) that characterizes the geometric phase associated to their argument (\ref{eq:weakvalueArgumentWithipreime}). Advantageously, all operators and states of two level systems are linked to a unit vector on the Bloch sphere. We note the initial and final states $\vec{i}$ and $\vec{f}$. Without loss of generality, we consider an observable of the form (\ref{eq:generalObservableAdefinition}), written as 
\begin{equation}
\hat{O_r}=a\left(\hat{I}+\gamma\ \vec{r}\cdot\hat{\vec{\sigma}}\right),
\end{equation}
where $\vec{r}$ is a unit vector and $a$ and $\gamma$ are real constants. Later on, $\gamma$ will appear as the relevant parameter for studying geometric phases. When $\gamma=0$, the weakly measured observable is the identity. When $\gamma=1$, it is proportional to a projector. Then, when $\gamma\rightarrow\infty$, it tends to a linear combination of the Pauli matrices, proportional to $\vec{r}\cdot\hat{\vec{\sigma}}$. From (\ref{eq:expression_weak_value_general_observable}) and (\ref{eq:argument_weak_value_general_observable}), we readily obtain the weak value and its argument: 
\begin{eqnarray}
\label{eq:weak_value_observable_two_level_systems}
&&O_{r,w}=a\frac{1+\vec{f}\cdot\vec{i}+\gamma\left(\vec{f}\cdot\vec{r}+\vec{r}\cdot\vec{i}\right)+\textrm{i}\gamma\,\vec{f}\cdot\left(\vec{r}\times\vec{i}\right)}{1+\vec{f}\cdot\vec{i}},
\\
\label{eq:argument_weak_value_observable_two_level_system}
&&\arg O_{r,w}=\arctan\frac{\gamma\,\vec{f}\cdot\left(\vec{r}\times\vec{i}\right)}{1+\vec{f}\cdot\vec{i}+\gamma\left(\vec{f}\cdot\vec{r}+\vec{r}\cdot\vec{i}\right)}+\phi\left(O_{r,w}\right),
\end{eqnarray}
with $\phi\left(O_{r,w}\right)$ giving the appropriate quadrant for the argument, on the basis of the signs of the real and imaginary parts of the weak value (\ref{eq:expression_phi}). It should be highlighted that the argument of the weak value depends chiefly on $\gamma$. The parameter $a$ only contributes a 0 or $\pi$ term through $\phi\left(O_{r,w}\right)$, depending on its sign. Interestingly, the parameter $\gamma$ plays a role similar to a measurement strength, from which the geometric phase (\ref{eq:argument_weak_value_observable_two_level_system}) emerges \cite{cho2019NatPhysGMphaseWM}.

The projector  $\hat{\Pi}_{i'}=\hat{O_r}\hat{\Pi}_i\hat{O_r}/ \Tr\hat{\Pi}_i\hat{O}_r^2$ (\ref{eq:iprimeProjectorDefinition}) connects the geometric phase (\ref{eq:argument_weak_value_observable_two_level_system}) to the argument of a Bargmann invariant. Its Bloch sphere vector is
\begin{eqnarray}
\label{eq:expression_ip_two_level_systems}
\vec{i'}=\frac{1}{1+2\gamma\,\vec{r}\cdot\vec{i}+\gamma^2}\left[\left(1-\gamma^2\right)\vec{i}+2\gamma\left(1+ \gamma\,\vec{r}\cdot\vec{i}\right)\vec{r}\right].
\end{eqnarray}
Figure \ref{fig:i_prime-prime} depicts the evolution of the $\vec{i}'$ vector as a function of the observable parameter $\gamma$. When $\gamma=0$,  $\vec{i'}$ is the pre-selected state $\vec{i}$ because the observable is the identity. With $\gamma=1$, $\vec{i'}$ corresponds to $\vec{r}$ since for that value of $\gamma$, the operator itself is already a projector. When $\gamma\rightarrow\infty$, the observable is proportional to the Pauli operator $\vec{r}\cdot\hat{\vec{\sigma}}$. Then, $\vec{i'}$ is the mirror image $\vec{i}_m$ of the initial vector $\vec{i}$ with respect to the direction $\vec{r}$ \cite{cormann2016revealing}:
\begin{equation}
    \vec{i}_m=-\vec{i}+2\left(\vec{i}\cdot\vec{r}\right)\vec{r}.
\end{equation}
All the possible locations of $\vec{i}'$ form the great circle that connects the initial state $\vec{i}$ to the direction $\vec{r}$ associated to the operator. Positive values of $\gamma$ correspond to the arc linking $\vec{i}\rightarrow\vec{r}\rightarrow\vec{i}_m$, while the negative values of gamma give the complementary arc $\vec{i}\rightarrow -\vec{r}\rightarrow\vec{i}_m$. Knowing $\vec{i}^\prime$, it becomes possible to represent geometrically the argument of the weak value on the Bloch sphere, as a function of $\gamma$ for fixed pre- and post-selected states, in a manner similar to figure \ref{fig:ray_space_triangle} (a). This is the main appeal of this projector.

\begin{figure}[tb]
 \centering
    \includegraphics[width=0.46\textwidth]{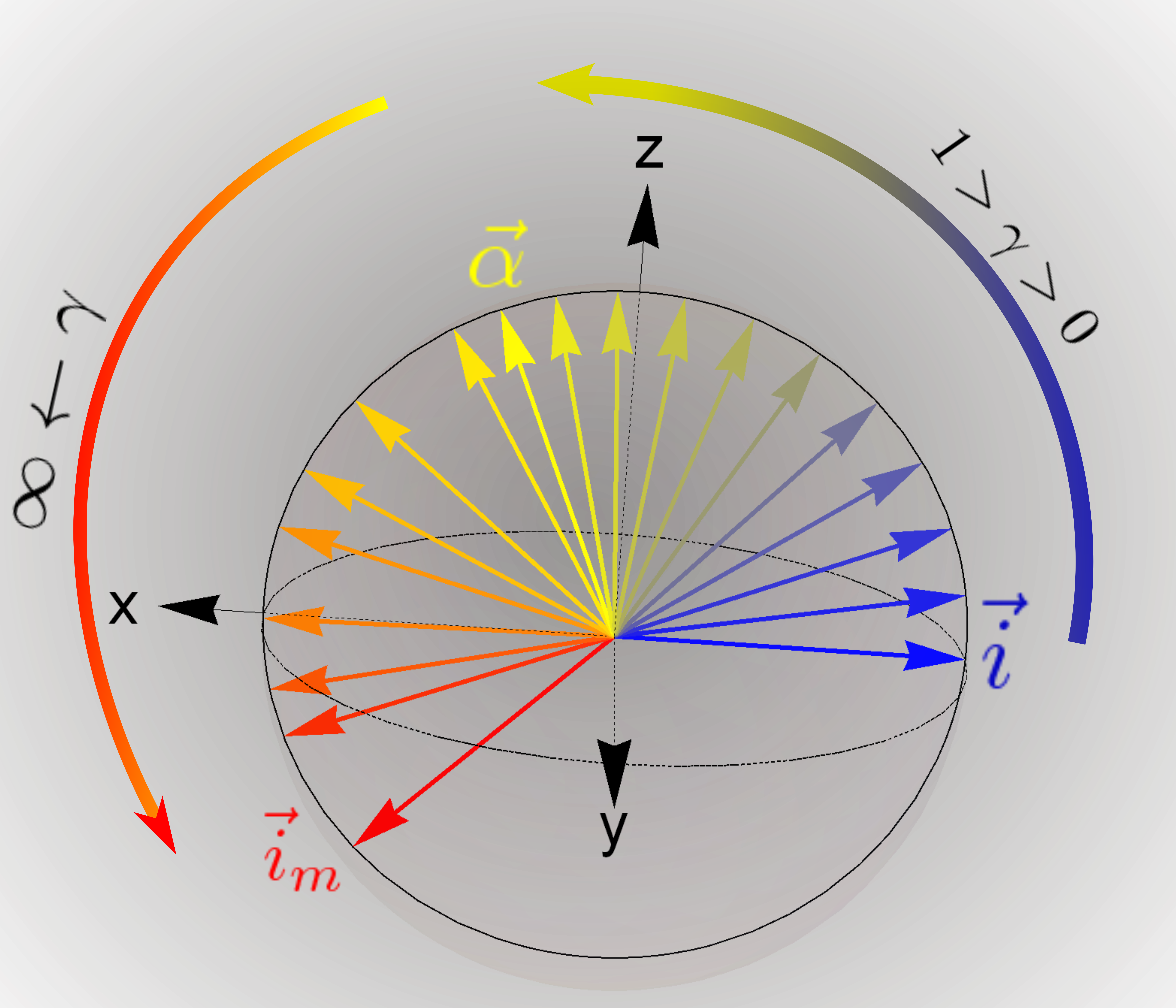}
\caption{Representation of the vector $\vec{i'}$ on the Bloch sphere for positive values of the parameter $\gamma$. The blue vector is the initial state, the yellow vector is the vector $\vec{r}$ associated to the observable. The red vector $\vec{i}_m$ is the mirror image of the initial state with respect to the vector $\vec{r}$.} \label{fig:i_prime-prime}
\end{figure}
At a more quantitative level, figure \ref{fig:angles} represents the smallest angles existing between the various relevant vectors. Two angles are constant: $\phi_{i r}$ between the initial state $\vec i$ and the operator vector $\vec r$, as well as $\phi_{i i_m}$ between the initial state and its mirror image $\vec{i}_m$ through $\vec{r}$. The evolution of the angle $\phi_{i i^\prime}$ between the initial state $\vec{i}$ and the projector $\vec{i}^\prime$ goes from 0 for $\gamma=0$ (when the operator is the identity), to $\phi_{i r}$ for $\gamma=1$ (when the operator $\hat{O}_r$ is the projector on $\vec{r}$), to the maximum value of $\pi$ for $\gamma=-1/\cos \phi_{i r}$. The latter corresponds to $\hat{O}_r\ket{\psi_i}=0$. In this case, both the weak value $O_{r,w}$ and the average value $\langle O_r \rangle_{\psi_i}$ are nul, and the argument of the weak value is undefined. This value of $\gamma$ delimits the parameter ranges for which the average value contributes a factor 0 or $\pi$ to the geometric phase (\ref{eq:weakvalueArgumentWithipreime}), according to its sign. Beyond this critical value of $\gamma$, the value of the $\phi_{i i^\prime}$ angle decreases and tends to $\phi_{i i_m}$. We also see that the angle $\phi_{r i^\prime}$ between $\vec r$ and $\vec{i}^\prime$ is equal to $\phi_{i r}$ when $\gamma=0$ or $\gamma\rightarrow\infty$, as these limiting cases correspond to mirror images with respect to $\vec r$. When the operator is a projector ($\gamma=1$), $\phi_{r i^\prime}=0$. Finally, we observe that $\phi_{i i^\prime}$ parametrizes the longitude along the great circle arc described by  $\vec{i}^\prime$. $\phi_{i i^\prime}$ can be expressed solely in terms of $\phi_{i r}$ and $\gamma$ by projecting (\ref{eq:expression_ip_two_level_systems}) on the initial state.

\begin{figure}[tb]
 \centering
    \includegraphics[width=0.75\textwidth]{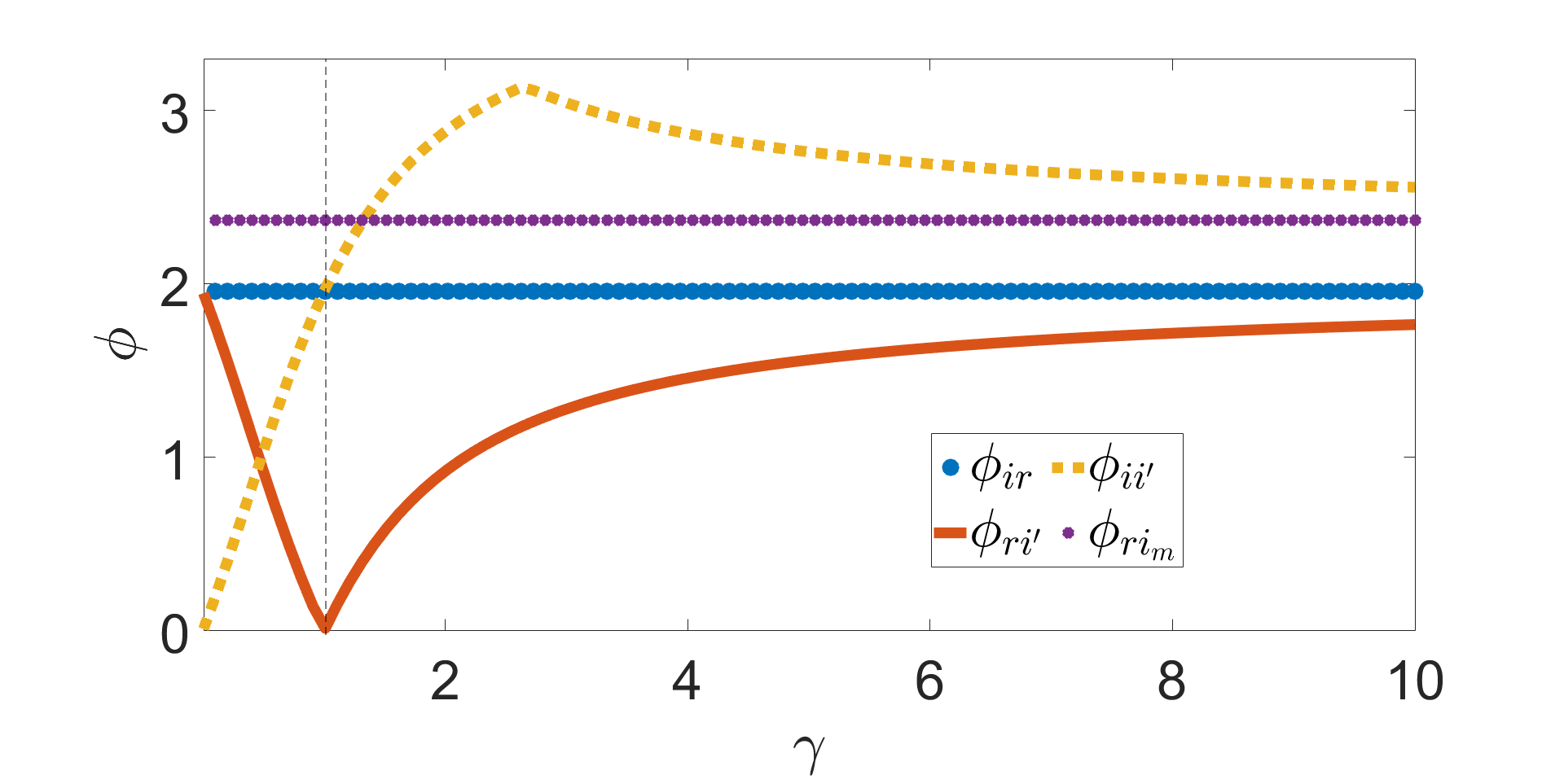}
\caption{Evolution of the smallest angles subtended by the four vectors $\vec{i}$ (arbitrary initial state), $\vec{r}$ (vector associated to the arbitrary qubit operator $\hat{O}_r$), $\vec{i}_m$ (mirror image of $\vec{i}$ through the direction $\vec{r}$) and $\vec{i}^\prime$ (Bloch sphere direction giving the geometric phase associated to the weak value as half a solid angle). In yellow: the angle $\phi_{i i^\prime}$ between $\vec{i}$ and $\vec{i}^\prime$. In dark orange: the angle $\phi_{r i^\prime}$ between $\vec{r}$ and $\vec{i}^\prime$. In blue and purple: the constant angles $\phi_{i r}$ and $\phi_{r i_m}= 2 (\pi-\phi_{i r})$, between $\vec{i}$ and $\vec{r}$ and between $\vec{r}$ and $\vec{i}_m$, respectively.  } \label{fig:angles} 
\end{figure}

\section{Projectors on degenerate subspaces and Hermitian quantum gates}
Sections \ref{section:projC3} and \ref{section:projCN} dealt with the weak values of projectors on pure states. However, projectors on degenerate subspaces also arise in practice. For example, the square of a spin-1 operator is associated to a doubly degenerate subspace, which plays an essential role in proofs of quantum contextuality \cite{peres1991JPHYSAcontextuality33rays}. As shown in \ref{appendix_B}, an arbitrary projector $\hat{P}$ on a $k$-degenerate subspace of $\mathbb{C}^n$  takes the form
\begin{equation}\label{eq:genProjectorDefDef}
\hat{P}=\frac{k}{N}\hat{I}_N+\sqrt{\frac{k(N-k)}{2N}} \vec\rho\cdot\vec{L},
\end{equation}
where the normalized, real vector $\vec{\rho}$ on the $S^{N^2-2}$ sphere is constrained by the star product according to
\begin{equation}\label{eq:genProjectorDefstarProduct}
\vec\rho\star\vec{\rho}=\frac{N-2k}{N-2}\sqrt{\frac{N-1}{k(N-k)}}\vec{\rho}.
\end{equation}
Setting $k=1$ in (\ref{eq:genProjectorDefDef}) and (\ref{eq:genProjectorDefstarProduct}), we recover the projectors on pure states. Then, $\vec\rho$ is associated to a quantum state in  $\mathbb{C}\textrm{P}^{N-1}$, but not otherwise (see example in \ref{appendix_B}).

The specific expressions of the weak value and its argument can be deduced as a straightforward application of (\ref{eq:expression_weak_value_general_observable}) and (\ref{eq:argument_weak_value_general_observable}). The point we actually wish to stress here is that, for such observables, the projector $\hat{\Pi}_{i^\prime}$ (\ref{eq:weakvalueArgumentWithipreime}) of $\mathbb{C}\textrm{P}^{N-1}$, whose argument of the Bargmann invariant with the pre- and post-selected states is linked to the weak value geometric phase, has a straightforward interpretation: it is simply the projection of the initial state on the degenerate subspace covered by $\hat{P}$. Its state vector is given by  $\hat{P}\ket{\psi_{i}}=\ket{\psi_{i^\prime}}$. Therefore, the argument of the weak value $P_w$ equals half of the symplectic area of the geodesic triangle with vertices $\ket{\psi_f}$, $\ket{\psi_i}$ and $\ket{\psi_{i^\prime}}$. Indeed, the average value a projector is always positive $\bra{\psi_i}\hat{P}\ket{\psi_i}\ge 0$ and, thus, dos not contribute to the geometric phase (\ref{eq:weakvalueArgumentWithipreime}). 

Another class of related observables gives rise to a nicely geometric interpretation of the projector $\hat\Pi_{i^\prime}$ present in the effective Bargmann invariant. It comprises all the observables that are both unitary and Hermitian. Amongst them, we recover many multi-qubit quantum gates, such as the the CNOT, CZ, SWAP, Toffoli and CSWAP gates, as well as, obviously, the Hadamard gate and all the Pauli gates acting on single qubits. Interestingly all the Hermitian unitary operators are connected to the projectors defined herebefore (\ref{eq:genProjectorDefDef}). Indeed from, any projector, we can build an observable $\hat{S}=2 \hat{P} - \hat{I}_N$ that is both Hermitian and unitary. Therefore,
\begin{equation}\label{eq:genHemtianUnitaryrDef}
\hat{S}=\frac{2k-N}{N}\hat{I}_N+\sqrt{\frac{2k(N-k)}{N}} \vec\rho\cdot\vec{L},
\end{equation}
where the unit vector $\vec\rho$ must obey the star product condition (\ref{eq:genProjectorDefstarProduct}) since it is linked to a $k$-degenerate projector $\hat{P}$. The state produced by $(2 \hat{P} - \hat{I}_N)\ket{\psi_i}=\ket{\psi_{i^\prime}}$ results from a generalized reflection of the complex state vector $\ket{\psi_i}$ in $\mathbb{C}^N$ with respect to the subspace corresponding to $\hat{P}$. This provides us with a quite elegant interpretation of the state contributing to the effective Bargmann invariant, reminiscent of the role played by the mirror image of the initial vector on the Bloch sphere in the case of Pauli operators (see figure \ref{fig:i_prime-prime}). For example, in qutrit systems, such an observable $\hat{S}$ would fundamentally flip the sign of one component of the state vector when it is expressed in the diagonal representation of the related projector $\hat{P}$. Therefore, the vector $\vec{i}^\prime$ in $S^7$ would follow from a peculiar reflection symmetry flipping the sign of the initial state vector components in four of the eight dimensions (in \ref{appendix_E}, see the expression \ref{eq:stateS7} of an arbitrary state on $S^7$ given as a function of its Hilbert space representation, where one should change the sign of a component $n_i$). Additionally, since the observable is unitary, the geodesic arc  connecting the states $\ket{\psi_i}$ and $\ket{\psi_{i^\prime}}$ in $\mathbb{C}\textrm{P}^{n-1}$ could correspond to an actual progressive evolution of the system. 

\section{Beyond weak measurements: average values and quantum uncertainties}\label{section:uncertainties}
Finally, we point out that we recover useful quantum mechanical expressions pertaining to average values of observables by setting identical initial and final states in the weak value  (\ref{eq:expression_weak_value_general_observable}). It is also helpful to look at expressions involving the variance in order to see how their definitions involve the star $\star$ and wedge $\wedge$ products of SU($N$).

We suppose that the quantum system is in the state $\ket{\psi}$, characterized by the vector $\vec{i}=\vec{i}\star\vec{i}$. We consider two general observables $\hat{A}=a_I \hat{I}_N + a_L\ \vec\alpha\cdot\hat{\vec{L}}$ and  $\hat{B}=b_I \hat{I}_N + b_L\ \vec\beta\cdot\hat{\vec{L}}$, where we use greek letters for the operator vectors as they do not necessarily correspond to a state on the $S^{N^2-2}$ sphere. Then the average value is related to the 8-dimensional Euclidean scalar product between the two vectors on the sphere:
\begin{equation}\label{eq:averageValueOnSphere}
\av{A}=\bra{\psi}\hat{A}\ket{\psi}=a_I+ a_L \sqrt{2\frac{N-1}{N}}\, \vec{i}\cdot\vec{\alpha},
\end{equation}
where we set $\vec{f}=\vec{i}$ in (\ref{eq:expression_weak_value_general_observable}). In \ref{appendix_C}, we compute the particular expressions of the squared operator $\hat{A}^2$, the commutator $[\hat{A}, \hat{B}]$ and the anticommutator $\{\hat{A}, \hat{B}\}$. These operators appear in the Heisenberg uncertainty relations as
\begin{equation}\label{eq:HeisenbergUncertainty}
\textrm{Var}(A) \textrm{Var}(B)-\textrm{Cov}^2(A,B)\ge \frac{1}{4}|\av{[\hat{A},\hat{B}]}|^2,
\end{equation}
where the variance is defined as usually by $\textrm{Var}(A)=\av{A^2}-\av{A}^2$ and the (symmetric) covariance by $\textrm{Cov}(A,B)=\frac{1}{2}\av{\{\hat{A},\hat{B}\}}-\av{A}\av{B}$. using these definitions, we find
\begin{eqnarray}
&&\textrm{Var}(A)=\frac{2 }{N}a_L^2[1-(N-1)(\vec\alpha\cdot\vec{i}\,)^2+(N-2)\ \vec\alpha\star\vec\alpha\cdot\vec{i}\,] \\
&& \textrm{Cov}(A,B)=\frac{2 }{N}a_L b_L[\vec{\alpha}\cdot\vec{\beta}-(N-1)(\vec\alpha\cdot\vec{i}\,)(\vec\beta\cdot\vec{i}\,)+(N-2)\ \vec\alpha\star\vec\beta\cdot\vec{i}\,] \\
&& \av{[\hat{A},\hat{B}]}=2 \textrm{i}  a_L b_L \sqrt{2\frac{N-1}{N}}\ \vec{\alpha}\wedge\vec{\beta}\cdot\vec{i}
\end{eqnarray}
Thus the two invariants that involve the wedge and star products that are present in the argument of weak values emerge as well in the Heisenberg uncertainty relationship. In particular, we observe that the numerator of the argument of the weak value, which involves the wedge product,  is therefore proportional to an average, in the initial state, of a commutator. The operators in this commutator are the weakly measured observable and the final state.  (Note that the roles of the initial and final state could be switched using a cyclic permutation of the three vectors.) Ultimately, these two invariants are built to provide  contribution involving the three vectors in a fully symmetric or anti-symmetric way. It is suggested in the literature that the average value of the square of the commutator is a predictor of quantum chaos \cite{maldacena2016JHEPquantumchaos}. This can also be evaluated to be
\begin{equation}
\av{|[\hat{A},\hat{B}]|^2}=\frac{8}{N} a_L b_L [\| \vec{\alpha}\wedge\vec{\beta}\|^2+(N-2) (\vec{\alpha}\wedge\vec{\beta})\star(\vec{\alpha}\wedge\vec{\beta})\cdot\vec{i}\,].
\end{equation}

\section{Conclusion}
We described on hyperspheres the geometrical properties of the weak values of general observables. For projectors on pure states of $N$-level systems, the argument of the weak value is the argument the Bargmann invariant of the initial, projector and final states. The argument of the Bargmann invariant and, hence, the argument of the weak value represent a geometric phase that is associated to half the symplectic area of the geodesic triangle in the projective space $\textrm{CP}^{N-1}$. The states are constrained to a $(2N-2)$-dimensional subspace of the unit sphere $S^{N^2-1}$, which generalizes the Bloch sphere. For all observables, we express the weak value and its argument in terms of  three Euclidean vectors located on $S^{N^2-1}$: formulas involve the standard Euclidean scalar product, as well as two vectorial operations inherited from SU($N$), represented by the star $\star$ and $\wedge$ products. We showed that the argument of the weak value is always related to a Bargmann invariant of three projectors, even when the observable probed by the weak measurement is not a projector. Thus we found a geometric depiction of the argument of any weak value in terms of a symplectic area of a geodesic triangle in complex projective space. For arbitrary observables of two-level systems, the symplectic area corresponds to the solid angle subtended by the three vectors associated to the initial state, the effective projector linked to the weak value, and the final state. We studied on the Bloch sphere how the projector associated to the effective Bargmann invariant evolves as a function of the weakly probed observable in a two-level system. We also investigated the geometric operations behind this projector in higher dimensional systems, for arbitrary projectors on degenerate subspaces and for Hermitian quantum gates. We produced the weak values of the generators of SU(N), including the Pauli and Gell-Mann matrices, which are essential to all spin and polarization applications of weak measurements. The formalism used here applies to both weak and average values, and we illustrated its usefulness beyond weak measurements by expressing the quantities intervening in the Heisenberg inequalities.  

Our geometric description applies to all weak measurements of discrete observables. Thus, it offer an extensive range of target applications. For example, it would especially benefit to studies of the three-dimensional polarization of light, of the orbital angular momentum of light beams used for quantum information processing tasks in large dimensions, of interferometers involving particles with spin larger than $\frac{1}{2}$, as well as of quantum paradoxes where the phase plays an important role. Indeed, as we focused our attention to the argument of the weak value, instead of the real part of the weak value (as is mostly done in the litterature), we provide particular insight into the interferometric aspects of weak values and weak measurements. There is an argument that the real part of the weak values contains the information about the system, while the imaginary part relates more to the meter \cite{dressel2012significance}. Our geometrical investigation of the argument of weak values
 advocates a holistic interpretation of the weak value. We particularly expect this approach to reveal its usefulness in fields in which the phase is essential, such as interferometry or quantum computing tasks.

\section*{Acknowledgments}
Y.C. is a Research Associate of the Fund for Scientific Research F.R.S.-FNRS. This research was supported by the Action de Recherche Concertée WeaM at the University of Namur (19/23-001).

\appendix
\section{Conventions for generators of SU($N$)} \label{appendix_A}
In this paper, as a convention, we have chosen to use the $N^2-1$ traceless Hermitian generators of $\textrm{SU}(N)$ that arise from the generalization of the Pauli and Gell-Mann matrices, thereafter noted $\hat{\vec{L}}$. From a strictly formal point of view, these operators would probably be best seen as twice the proper generators of $\textrm{SU}(N)$, typically noted $\hat{\vec{T}}$. However, we think this choice is natural in order to express quantum states on the Bloch sphere and generalized Bloch spheres in terms of standard operators. The generators usually defined by $\hat{\vec{T}}$ follow the properties
\begin{eqnarray}
\left[\hat{T}_a, \hat{T}_b\right]&=& \textrm{i} \sum_{c} f_{abc}\hat{T}_c, \\ \nonumber
\{\hat{T}_a,\hat{T}_b\}&=&\frac{1}{N}\delta_{ab}\hat{I}_N+\sum_{c} d_{abc}\hat{L}_c ,\\ \nonumber
\Tr\, \hat{T}_a\hat{T}_b&=&\frac{1}{2}\delta_{ab},\\ \nonumber
\hat{T}_a\hat{T}_b&=&\frac{1}{2N}\delta_{ab}\hat{I}_N+\frac{1}{2}\sum_{c}\left(d_{abc}+\textrm{i} f_{abc}\right)\hat{T}_c.
\end{eqnarray}
Even though these generators are mathematically convenient, quantum physics makes extensive use of the Pauli matrices $\hat{\vec\sigma}$. Their SU(3) counterparts are the well-known Gell-Mann matrices $\hat{\vec\lambda}$. They share the following properties with their $\textrm{SU}(N)$ generalization $\hat{\vec{L}}$:
\begin{eqnarray}
\label{eq:properties_L}
\left[\hat{L}_a, \hat{L}_b\right]&=& 2 \textrm{i} \sum_{c} f_{abc}\hat{L}_c, \\ \nonumber
\{\hat{L}_a,\hat{L}_b\}&=&\frac{4}{N}\delta_{ab}\hat{I}_N+2 \sum_{c} d_{abc}\hat{L}_c, \\ \nonumber
\Tr\,\hat{L}_a\hat{L}_b&=&2\delta_{ab},\\ \nonumber
\hat{L}_a\hat{L}_b&=&\frac{2}{N}\delta_{ab}\hat{I}_N+\sum_{c}\left(d_{abc}+ \textrm{i} f_{abc}\right)\hat{L}_c.
\end{eqnarray}
The anti-symmetric structure constants and the symmetric constants of $\textrm{SU}(N)$ are connected to the generators using
\begin{eqnarray}
f_{abc}&=&-\frac{1}{4}\textrm{i}\Tr(\hat{L}_a [\hat{L}_b,\hat{L}_c])=-2\textrm{i}\Tr(\hat{T}_a [\hat{T}_b,\hat{T}_c]), \\
d_{abc}&=&\frac{1}{4}\Tr (\hat{L}_a\{\hat{L}_b,\hat{L}_c\})= 2 \Tr(\hat{L}_a\{\hat{L}_b,\hat{L}_c\}),
\end{eqnarray}
so that $\hat{\vec{L}}=2\, \hat{\vec{T}}$. This simple proportionality relationship provides the conversion rule between the two conventions, should anyone wish to use expressions with the $\hat{\vec{T}}$ generators. From now on, we will exclusively work with the $\hat{\vec{L}}$ generators.

\section{Conventions for star product and projectors} \label{appendix_B}
From the symmetric constants $d_{abc}$ of SU($N$), we can construct a symmetric product called the star product. Given two $(N^2-1)$-dimensional vectors, the $\star$ product produces a vector with components $(\vec\alpha\star\vec\beta)_c=c_s \sum_{ab}{d_{abc}\alpha_a \beta_b}$, where $c_s$ is a proportionality constant. It would be convenient to simply set $c_s=1$ (as done for the definition of the anti-symmetric wedge product $(\vec\alpha\wedge\vec\beta)_c= \sum_{ab}{f_{abc}\alpha_a \beta_b}$, which is built on the structure constants). However, the original definition of the star product in the literature \cite{mallesh1997generalized, khanna1997AnnPhysGMphaseSU3} used another convention: the star product identified a proper quantum state from $\mathbb{C}P^{2}$ on the $S^{7}$ sphere. In the following, we explain thus how we generalized the star product from SU(3) to SU($N$), based on this earlier choice. We also explain our normalization convention for projectors, as they are  related.

A general projector $\hat{P}$ in $\mathbb{C}^N\times\mathbb{C}^N$ acting on states in  $\mathbb{C}^N$ is defined by the relation $\hat{P}^2=\hat{P}$. In addition, its trace is an integer number that is lower or equal to the dimension $N$: $\Tr \hat{P}=k$, with $1\leq k\leq N$ representing the dimension of the projector subspace (essentially the degeneracy of the eigenvalue $1$). We pose $\hat{P}=\frac{k}{N} \hat{I}_N+ c_p \, \vec{\beta}\cdot\hat{\vec{L}}$ to meet the trace condition. The positive constant 
\begin{equation}\label{eq:defcp}
c_p=\sqrt{\frac{k(N-k)}{2N}}
\end{equation}
ensures that $\vec{\beta}$ is always a normalized vector ($\vec{\beta}\cdot\vec{\beta}=1$) on the $S^{N^2-2}$ unit sphere. Its value originates from the projector condition
\begin{equation}\label{eq:ProjCondAnnex}
\hat{P}^2=(\frac{k^2}{N^2}+\frac{2}{N}c_p^2)\hat{I}_N+\frac{2k}{N} c_p\ \vec{\beta}\cdot\hat{\vec{L}}+c_p^2 \sum_{abc}{d_{abc}\beta_a \beta_b \hat{L}_c}=\frac{k}{N} \hat{I}_N+ c_p\, \vec{\beta}\cdot\hat{\vec{L}},
\end{equation}
where we used (\ref{eq:properties_L}) to expand the SU($N$) generator square $(\vec{n}\cdot\hat{\vec{L}})^2$. Additionaly, (\ref{eq:ProjCondAnnex}) constraints the vector $\vec{\beta}$ through the star product 
\begin{equation}\label{eq:constraint_starproduct}
\frac{1}{c_s}(\vec\beta\star\vec\beta)_c=\sum_{ab}{d_{abc}\beta_a \beta_b}=(1-\frac{2k}{N})\frac{1}{c_p}\beta_c,
\end{equation}
In the literature \cite{mallesh1997generalized, khanna1997AnnPhysGMphaseSU3}, the star product was defined in SU(3) by imposing that this condition becomes $\vec\beta\star\vec\beta=\vec\beta$ for projectors on pure states (case $k=1$). Therefore, imposing $\vec\beta\star\vec\beta=\vec\beta$ for pure states, we find the value of the constant $c_s$ in SU($N$):
\begin{equation}\label{eq:defconstantcs}
c_s=\frac{N c_p}{N-2}=\frac{1}{N-2}\sqrt{\frac{N(N-1)}{2}},
\end{equation}
which thus defines the star product in SU($N$) \cite{Byrd2003PRAstardimN}. As a result, an arbitrary projector takes the form
\begin{equation}
\hat{P}=\frac{k}{N} \hat{I}_N+ \sqrt{\frac{k(N-k)}{2N}} \, \vec{\beta}\cdot\hat{\vec{L}},
\end{equation}
with the accompanying star product constraint resulting from (\ref{eq:constraint_starproduct}):
\begin{equation}\label{eq:def_starproductconstraintAllProj}
\vec{\beta}\star\vec{\beta}=\frac{N-2k}{N-2}\sqrt{\frac{N-1}{k(N-k)}} \vec{\beta}.
\end{equation}
Note that, only when $k=1$, does this projector correspond to a quantum state from $\mathbb{C}\textrm{P}^{n-1}$. For example, a projector on an $(N-1)$-dimensional subspace obeys $\vec\beta\star\vec\beta=-\vec\beta$; it is the opposite of the vector associated to the projection on the complementary 1-dimensional subspace (this one is a quantum state).

In summary, our convention sets the constant $c_p$ defining an arbitrary  projector so that we always work with vectors belonging to hyperspheres of unit radius when using the operators $\hat{\vec{L}}$ as generators of SU($N$) \cite{goyal2016JPhysABlochQudits, mallesh1997generalized, khanna1997AnnPhysGMphaseSU3, byrd1998JMPHYSsu3starp, Byrd2003PRAstardimN}. Then, in order to define the star product, we follow the literature convention that the set of vectors representing pure quantum states is equivalent to the vectors invariant under the star product (in the sense of $\vec{r}\star\vec{r}=\vec{r}$) \cite{goyal2016JPhysABlochQudits, mallesh1997generalized, khanna1997AnnPhysGMphaseSU3, byrd1998JMPHYSsu3starp, Byrd2003PRAstardimN}. We note that, when working with generalized Bloch spheres, some authors prefer to set $c_p=1$ and deal with unnormalized vectors \cite{kimura2003PLAblochnlevel, Bertlmann2008JPhysAblochqudits}. This would be inconvenient for us, as many expressions linked to weak values are invariant under permutations of the related vectors, some of which would be normalized and others not. Managing the vector normalization status complicates geometric descriptions as well. On the other hand, if the star product were defined initially with the constant $c_s=1$, the constant $c_s$ would not appear when the product of two generators is expressed in terms of the star product, such as in the weak value formula  (\ref{eq:WV-real}). Were it the case, the projector condition for pure state would have been $\vec{r}\star\vec{r}=\frac{1}{c_s}\vec{r}$. A few authors working with unnormalized vectors chose to redefine to the star product with $c_s=1$ \cite{Siennicki2001AnnPhysStarWedge, graf2021PRBstarpcrossp}.

\section{Computation of the weak value and Bargmann invariant}\label{appendix_C}

Computing weak values involves the traces of products of two and three operators. Considering three arbitrary generators $\vec\alpha\cdot\hat{\vec{L}}$, $\vec\beta\cdot\hat{\vec{L}}$ and $\vec\gamma\cdot\hat{\vec{L}}$, using (\ref{eq:properties_L}), we find the products and traces
\begin{eqnarray}
\label{eq:product2generators}
(\vec\alpha\cdot\hat{\vec{L}})(\vec\beta\cdot\hat{\vec{L}})=\frac{2}{N}\vec\alpha\cdot\vec\beta\ \hat{I}_N+\frac{1}{c_s} (\vec\alpha\star\vec\beta)\cdot\hat{\vec{L}}+\textrm{i} \ (\vec\alpha\wedge\vec\beta)\cdot\hat{\vec{L}}, \\
\label{eq:product2generators-trace}
\Tr[(\vec\alpha\cdot\hat{\vec{L}})(\vec\beta\cdot\hat{\vec{L}})]=2\ \vec\alpha\cdot\vec\beta,
\end{eqnarray}
where $c_s$ is defined in (\ref{eq:defconstantcs}). For three generators, we have
\begin{eqnarray}
\label{eq:product3generators}
(\vec\alpha\cdot\hat{\vec{L}})(\vec\beta\cdot\hat{\vec{L}})(\vec\gamma\cdot\hat{\vec{L}})=\frac{2}{N}\left[\frac{1}{c_s}(\vec\alpha\star\vec\beta)\cdot\vec\gamma+\textrm{i}(\vec\alpha\wedge\vec\beta)\cdot\vec\gamma\right] \hat{I}_N  \nonumber \\
\quad\quad\quad+\frac{2}{N}\vec\alpha\cdot\vec\beta\ (\vec\gamma\cdot\hat{\vec{L}})
+\frac{1}{c_s^2} [(\vec\alpha\star\vec\beta)\star\vec{\gamma}]\cdot\hat{\vec{L}}- \ [(\vec\alpha\wedge\vec\beta)\wedge\vec\gamma]\cdot\hat{\vec{L}} \nonumber\\
\quad\quad\quad+\textrm{i}\frac{1}{c_s} \left\{ [(\vec\alpha\wedge\vec\beta)\star\vec\gamma]\cdot\hat{\vec{L}}
+ [(\vec\alpha\star\vec\beta)\wedge\vec{\gamma}]\cdot\hat{\vec{L}}\right\}, \\
\label{eq:product3generators-trace}
\Tr [(\vec\alpha\cdot\hat{\vec{L}})(\vec\beta\cdot\hat{\vec{L}})(\vec\gamma\cdot\hat{\vec{L}})]=\frac{2}{c_s}(\vec\alpha\star\vec\beta)\cdot\vec\gamma+2\textrm{i}\ (\vec\alpha\wedge\vec\beta)\cdot\vec\gamma.
\end{eqnarray}
As the trace is invariant under unitary transformations, we see that the two quantities $(\vec\alpha\star\vec\beta)\cdot\vec\gamma$ and $(\vec\alpha\wedge\vec\beta)\cdot\vec\gamma$ present in (\ref{eq:product3generators-trace}) are also invariant under unitary transformations. From the properties of the $d_{abc}$ and $f_{abc}$ constants of SU($N$), the former is fully symmetric under permutations, while the latter is antisymmetric and changes sign under permutation of two vectors.

Now we define two projectors on pure states $\hat\Pi_i=\frac{1}{N}\hat{I}_N+c_p\ \vec{i}\cdot\hat{\vec{L}}$ and  $\hat\Pi_f=\frac{1}{N}\hat{I}_N+c_p\ \vec{f}\cdot\hat{\vec{L}}$ (with $c_p$ given by (\ref{eq:defcp}) with $k=1$), as well as two arbitrary Hermitian operators $\hat{A}=a_I \hat{I}_N + a_L\ \vec{\alpha}\cdot\hat{\vec{L}}$ and $\hat{B}=b_I \hat{I}_N + b_L\ \vec{\beta}\cdot\hat{\vec{L}}$. Following (\ref{eq:product2generators}), the product of the operators $\hat{A}$ and $\hat{B}$ becomes
\begin{eqnarray}
\hat{A}\hat{B}&=&a_I b_I \hat{I}_N+ a_L b_I\ \vec{\alpha}\cdot\hat{\vec{L}}+a_I b_L\ \vec{\beta}\cdot\hat{\vec{L}} + a_L b_L (\vec\alpha\cdot\hat{\vec{L}})(\vec\beta\cdot\hat{\vec{L}}) \nonumber \\
&=&(a_I b_I + \frac{2}{N}a_L b_L\ \vec\alpha\cdot\vec\beta) \hat{I}_N+ a_L b_I\ \vec{\alpha}\cdot\hat{\vec{L}}+a_I b_L\ \vec{\beta}\cdot\hat{\vec{L}} \nonumber \\ \label{eq:product2operatorsAnnex}
&+& a_L b_L \frac{1}{c_s}\ (\vec\alpha\star\vec\beta)\cdot\hat{\vec{L}}+  \textrm{i} a_L b_L\ (\vec\alpha\wedge\vec\beta)\cdot\hat{\vec{L}}.
\end{eqnarray}
When $\hat{A}=\hat{B}$, the latter simplifies to
\begin{equation}\label{eq:squareA-annex}
\hat{A}^2=(a_I^2 + \frac{2}{N} a_L^2) \hat{I}_N+ 2 a_I a_L\ \vec{\alpha}\cdot\hat{\vec{L}}+ a_L^2 \frac{1}{c_s}\ (\vec\alpha\star\vec\alpha)\cdot\hat{\vec{L}}.
\end{equation}
These expressions also allow us to compute the commutator and anti-commutator
\begin{eqnarray}
\label{eq:commutAB-Annex}
[\hat{A},\hat{B}]&=&  2\textrm{i} a_L b_L\ (\vec\alpha\wedge\vec\beta)\cdot\hat{\vec{L}}\\
\{\hat{A},\hat{B}\}&=& 2(a_I b_I + \frac{2}{N}a_L b_L\ \vec\alpha\cdot\vec\beta) \hat{I}_N+ 2a_L b_I\ \vec{\alpha}\cdot\hat{\vec{L}}+2a_I b_L\ \vec{\beta}\cdot\hat{\vec{L}} \nonumber \\ 
&+& 2 a_L b_L \frac{1}{c_s}\ (\vec\alpha\star\vec\beta)\cdot\hat{\vec{L}}
\end{eqnarray}
The latter were used in section \ref{section:uncertainties} to compute the variance and covariance of operators. Formula (\ref{eq:product2operatorsAnnex}) gives the product $\hat{\Pi}_f \hat{\Pi}_i$ of the two projectors, by simply setting $a_I=b_I=\frac{1}{N}$ and $a_L=b_L=c_p$, so that the trace of two projectors is
\begin{eqnarray}\label{eq:trace2proj}
\Tr\, (\hat{\Pi}_f \hat{\Pi}_i)= \frac{1}{N}+ 2 c_p^2\ \vec{f}\cdot\vec{i}= \frac{1}{N}[1+ (N-1)\ \vec{f}\cdot\vec{i}\,].
\end{eqnarray}
This is the denominator of the weak value (\ref{eq:weak_value_definition}). To obtain its numerator, we evaluate
\begin{eqnarray}
\hat{\Pi}_f \hat{A} \hat{\Pi}_i&=&\frac{a_I}{N^2} \hat{I}_N + \frac{a_I}{N}c_p\ \vec{i}\cdot\hat{\vec{L}} + \frac{a_L}{N^2}\ \vec{\alpha}\cdot\hat{\vec{L}}+\frac{a_L}{N}c_p (\vec{\alpha}\cdot\hat{\vec{L}})(\vec{i}\cdot\hat{\vec{L}}) \\ \nonumber
&+&\frac{a_I}{N} c_p\ \vec{f}\cdot\hat{\vec{L}} + a_I c_p^2 (\vec{f}\cdot\hat{\vec{L}})(\vec{i}\cdot\hat{\vec{L}})+\frac{a_L}{N}c_p (\vec{f}\cdot\hat{\vec{L}})(\vec{\alpha}\cdot\hat{\vec{L}}) \\ \nonumber
&+&a_L c_p^2 (\vec{f}\cdot\hat{\vec{L}})(\vec{\alpha}\cdot\hat{\vec{L}}) (\vec{i}\cdot\hat{\vec{L}}).
\end{eqnarray}
Then, the trace formulas (\ref{eq:product2generators-trace}) and (\ref{eq:product3generators-trace}) yield
\begin{eqnarray}\label{WV-Tr-denom-annex}
\Tr\, (\hat{\Pi}_f \hat{A} \hat{\Pi}_i)&=&\frac{a_I}{N} +2\frac{a_L}{N}c_p \ \vec{\alpha}\cdot\vec{i} 
 +2 a_I c_p^2 \ \vec{f}\cdot\vec{i}+2 \frac{a_L}{N}c_p \ \vec{f}\cdot\vec\alpha \\ \nonumber
&+&2 a_L c_p^2 \left[\frac{1}{c_s}(\vec{f}\star\vec\alpha)\cdot\vec{i}+\textrm{i}\ (\vec{f}\wedge\vec\alpha)\cdot\vec{i}\right].
\end{eqnarray}
The Bargmann invariant ensues from considering that $\hat{A}=\hat{\Pi}_r$ ($a_I=\frac{1}{N}$ and $a_L=c_p$),
\begin{eqnarray}
\Tr\, (\hat{\Pi}_f \hat{\Pi}_r \hat{\Pi}_i)&=&\frac{1}{N^2} +\frac{2}{N}c_p^2 (\vec{r}\cdot\vec{i} 
 + \vec{f}\cdot\vec{i}+\vec{f}\cdot\vec{r}) \\ \nonumber
&+&2c_p^3 \left[\frac{1}{c_s}(\vec{f}\star\vec{r})\cdot\vec{i}+\textrm{i}\ (\vec{f}\wedge\vec{r})\cdot\vec{i}\right],
\end{eqnarray}
while the weak value is simply given by the ratio of (\ref{WV-Tr-denom-annex}) with (\ref{eq:trace2proj}). The real and imaginary part of the weak value are thus
\begin{eqnarray}
\label{eq:WV-real}
\Re A_w &=& \frac{\frac{a_I}{N} +\frac{2 a_L}{N}c_p (\vec{\alpha}\cdot\vec{i}+ \vec{f}\cdot\vec\alpha) + 2 a_I c_p^2 \ \vec{f}\cdot\vec{i}+ 2 a_L c_p^2 \frac{1}{c_s}(\vec{f}\star\vec\alpha)\cdot\vec{i}} {\frac{1}{N}+ 2 c_p^2\ \vec{f}\cdot\vec{i}},\\
\label{eq:WV-im}
\Im A_w &=& \frac{ 2\textrm{i}\, a_L c_p^2 \ (\vec{f}\wedge\vec\alpha)\cdot\vec{i}} {\frac{1}{N}+ 2 c_p^2\ \vec{f}\cdot\vec{i}}.
\end{eqnarray}
The argument of the weak value is determined by $\arctan (\Im A_w/\Re A_w)$ while simultaneously taking the signs of (\ref{eq:WV-real}) and (\ref{eq:WV-im}) into account to recover the appropriate quadrant. The latter expressions allow for an easy conversion between possibly different conventions for the $c_p$ and $c_s$ constants, as discussed in \ref{appendix_B}.

\section{Properties of states on $S^{N^2-1}$ and the $\star$ and $\wedge$ products}\label{appendix_D}
In this section, we review a few properties of the state representation on the generalized Bloch sphere, in connection to the $\star$ and $\wedge$ products of SU($N$), as we believe this formalism is still unfamiliar to many. Our goal is to provide a glimpse on key aspects of the geometry and highlight information relevant to interpreting the various contributions to weak value formulas.

First of all, we consider the conditions defining an orthonormal basis of $\mathbb{C}\textrm{P}^{n-1}$: the projector orthogonality relationship $\hat{\vec\Pi}_i\hat{\vec\Pi}_j=\delta_{ij}\hat{\vec{\Pi}}_i$, as well as the resolution of the identity $\sum_{i=1}^N{\hat\Pi_{i}}=\hat{I}_N$. For orthogonal states, the former impose that the angles between the vectors are given by  $\vec{n}_i\cdot\vec{n}_j=-\frac{1}{N-1}$ (\ref{eq:trace2proj}). In addition, the later results in $\sum_i \vec{n}_i=0$. Thus, the vectors are all placed very symmetrically. For SU(2), these conditions show that vectors associated to orthogonal states are opposite, a well-known property of the standard Bloch sphere. For SU(3), the three vectors arising from a state basis all lie in a plane, with angles of 120° between them: their extremities form an equilateral triangle. For SU(4), the vectors build a tetrahedron. In larger dimensions, for SU(N), the arrangements remain extremely symmetric in a similar fashion, with the $N$ vectors residing in a subspace of $N-1$ dimensions. We see thus that orthogonal states do not correspond to orthogonal vectors in the Euclidean sense. Actually, an orthogonal vector $\vec{r}$ verifying $\vec{r}\cdot\vec{n}_i=0$ for all states of a given basis, obey $\Tr\, (\hat\Pi_r \hat\Pi_i)=\frac{1}{N}$  (\ref{eq:trace2proj}). It corresponds to a state with maximal relative uncertainty with respect to the measurement basis (such as between states belonging to two different mutually unbiased bases). Orthogonal quantum states verify the following two additional relationships for their associated vectors: their wedge product is nul $\vec{m}\wedge\vec{n}=0$ and their star product is located on the angle bissector of the two vectors  $\vec{m}\star\vec{n}=-\frac{1}{N-2}(\vec{m}+\vec{n})$. In the particular case of SU(3), the later results in the third basis vector being given by the star product of the other two ($\vec{n}_1\star\vec{n}_2=\vec{n}_3$).

The symmetric star product is not associative. The star product of an arbitrary normalized vector $\vec{\alpha}$ does not generally produce a normalized vector  ($\|\vec{\alpha}\star\vec{\alpha}\|\ne1$), with the exception of SU(3), where $\vec{\alpha}\star\vec{\alpha}$ always remain on $S^7$. Of course, all vectors associated to pure states also remain on $S^{N^2-1}$ in this manner, following the choice set by the definition of the $\star$ product. As the star product defines the condition for a vector to represent a state ($\vec{r}\star\vec{r}=\vec{r}$), we consider the particular case of the star product between two vectors associated to states. In that particular case, we have $(\vec{q}\star\vec{r}\,)\cdot\vec{q}=(\vec{q}\star\vec{r}\,)\cdot\vec{r}=\vec{q}\cdot\vec{r}$ (thanks to the fully symmetric nature of the product). Therefore, the star product of two projectors lies in the median hyperplane lying between $\vec{q}$ and ${\vec{r}}$, which is orthogonal to $\vec{q}-\vec{r}$, as could be expected from the symmetric properties of the product. In general, the star product of two vectors $\vec{q}$, $\vec{r}$ does not remain in the plane spanned by the two vectors (contrary to what we observed for two orthogonal states). Neither does it represent a state in general. We note that, operationally, an observable involving the vector $\vec{\alpha}\star\vec{\alpha}$ can be constructed from the square of an operator $(\vec\alpha\cdot\hat{\vec{L}})^2=\frac{2}{N}\hat{I}_N+\frac{1}{c_s} [(\vec{\alpha}\star\vec{\alpha})\cdot\hat{\vec{L}}]$  (\ref{eq:squareA-annex}), so that this vector contributes to quantum fluctuations (see section \ref{section:uncertainties}).

The anti-symmetric wedge product is not associative. It produces a vector orthogonal to the initial ones: $\vec{\alpha}\wedge\vec{\alpha}=0$ and, therefore, $\vec{\alpha}\wedge\vec{\beta}\cdot\vec{\alpha}=\vec{\alpha}\wedge\vec{\beta}\cdot\vec{\beta}=0$. However, due to the large number of dimensions involved, the wedge product selects an orthogonal direction amongst many available (contrary to the cross-product in three dimensions, for which the orthogonal direction to two non-parallel vectors is unique). The wedge product is intimately associated to the commutator (\ref{eq:commutAB-Annex}). From the Baker-Campbell-Hausdorff formula, the wedge product gives the unitary operator associated to the non-commutativity of consecutive unitary transformations, such as the generators of rotations: $e^{-\textrm{i}g \vec{\beta}\cdot\hat{\vec{L}}}e^{-\textrm{i}g \vec{\alpha}\cdot\hat{\vec{L}}}e^{\textrm{i}g \vec{\beta}\cdot\hat{\vec{L}}}e^{\textrm{i}g \vec{\alpha}\cdot\hat{\vec{L}}}\approx e^{\textrm{i}g^2 (\vec{\alpha}\wedge\vec\beta)\cdot\hat{\vec{L}}}$. From a practical point of view, this allows to construct an observable linked to the wedge product. This product gives thus also the direction of the effective transformation of an observable undergoing a small unitary transformation (from (\ref{eq:commutAB-Annex}) as $e^{\textrm{i} g\hat{A}}\hat{B}e^{-\textrm{i} g\hat{A}}\approx \hat{B}+\textrm{i} g[\hat{A},\hat{B}]$).

The wedge and star products between two vectors representing states are orthogonal in the following sense: $(\vec{q}\star\vec{r})\cdot(\vec{q}\wedge\vec{r})=0$ (actually, we also checked up to to SU(6) using brute force calculation with a computer algebra system that this is true for any two vectors). Other relationships connect the star an wedge products in the case of pure states. By imposing that a projector $\hat\Pi_r$ remains a projector after a unitary transformation with generator $\vec\alpha\cdot\hat{\vec{L}}$, to first order, we obtain $\vec{\alpha}\wedge\vec{r}=2 \vec{r}\star(\vec{\alpha}\wedge\vec{r})$ (and more complex relationships can be deduced from second-order contributions).

\section{$\mathbb{C}P^2$ representation on $S^7$}\label{appendix_E}

Considering an arbitrary state $\ket{\psi}$, the coordinates on the corresponding hypersphere are obtained by $\Tr \hat{\Pi}_\psi \hat{\vec{L}}$.
 On $S^7$, the state $\ket{\psi}=\left(n_1 e^{\mathrm{i}\chi_1}, n_2 e^{\mathrm{i}\chi_2}, n_3 e^{\mathrm{i}\chi_3}\right)^T$ would thus become the gauge-invariant vector
\begin{eqnarray}\label{eq:stateS7} \nonumber
[n_1 n_2 \cos{(\chi_1-\chi_2)}, -n_1 n_2 \sin{(\chi_1-\chi_2)}, \frac{1}{2}(n_1^2-n_2^2), n_1 n_3 \cos{(\chi_1-\chi_3)} ,\\ \nonumber
 -n_1 n_3 \sin{(\chi_1-\chi_3)}, n_2 n_3 \cos{(\chi_2-\chi_3)} ,-n_2 n_3 \sin{(\chi_2-\chi_3)}, \\ 
 \frac{1}{2 \sqrt{3}}(n_1^2+n_2^2-2n_3^2)]^T.
\end{eqnarray}
A closed geodesic $(0, \sin s, \cos s)^T$ ($s \in [0, \pi]$) appears therefore as a tilted circle of radius $\sqrt{3}/2$ on $S^7$ (it is not a great-circle):
\begin{equation}
[0,0,-\frac{\sqrt{3}}{4}(1-\cos{2s}),0,0,\frac{\sqrt{3}}{2}\sin{2s},0,-\frac{1}{4}(1+3 \cos{2s})]^T.
\end{equation}

 \section*{References}
\bibliographystyle{iopart-num-YC}
\bibliography{Journal_AbbreviationDB.bib, biblio}

\end{document}